\documentclass[dvips]{acta}
\usepackage{supertabular,lscape,epsfig}
\usepackage{amssymb}
\usepackage{amsmath}
\DeclareSymbolFont{ppa}{OT1}{ppl}{m}{it}
\DeclareMathSymbol{\vv}{\mathalpha}{ppa}{'166}

\newfont{\hb}{rphvb at 10pt}
\newfont{\hbo}{rphvbo at 10pt}
\newfont{\bitt}{rptmbi at 12pt}
\newfont{\bits}{rptmbi at 11pt}

\SetPages{219}{245}

\SetVol{62}{2012}

\begin{document}

\newcommand{\TabApp}[2]{\begin{center}\parbox[t]{#1}{\centerline{
  {\bf Appendix}}
  \vskip2mm
  \centerline{\small {\spaceskip 2pt plus 1pt minus 1pt T a b l e}
  \refstepcounter{table}\thetable}
  \vskip2mm
  \centerline{\footnotesize #2}}
  \vskip3mm
\end{center}}

\newcommand{\TabCapp}[2]{\begin{center}\parbox[t]{#1}{\centerline{
  \small {\spaceskip 2pt plus 1pt minus 1pt T a b l e}
  \refstepcounter{table}\thetable}
  \vskip2mm
  \centerline{\footnotesize #2}}
  \vskip3mm
\end{center}}

\newcommand{\TTabCap}[3]{\begin{center}\parbox[t]{#1}{\centerline{
  \small {\spaceskip 2pt plus 1pt minus 1pt T a b l e}
  \refstepcounter{table}\thetable}
  \vskip2mm
  \centerline{\footnotesize #2}
  \centerline{\footnotesize #3}}
  \vskip1mm
\end{center}}

\newcommand{\MakeTableApp}[4]{\begin{table}[p]\TabApp{#2}{#3}
  \begin{center} \TableFont \begin{tabular}{#1} #4 
  \end{tabular}\end{center}\end{table}}

\newcommand{\MakeTableSepp}[4]{\begin{table}[p]\TabCapp{#2}{#3}
  \begin{center} \TableFont \begin{tabular}{#1} #4 
  \end{tabular}\end{center}\end{table}}

\newcommand{\MakeTableee}[4]{\begin{table}[htb]\TabCapp{#2}{#3}
  \begin{center} \TableFont \begin{tabular}{#1} #4
  \end{tabular}\end{center}\end{table}}

\newcommand{\MakeTablee}[5]{\begin{table}[htb]\TTabCap{#2}{#3}{#4}
  \begin{center} \TableFont \begin{tabular}{#1} #5 
  \end{tabular}\end{center}\end{table}}

\newfont{\bb}{ptmbi8t at 12pt}
\newfont{\bbb}{cmbxti10}
\newfont{\bbbb}{cmbxti10 at 9pt}
\newcommand{\uprule}{\rule{0pt}{2.5ex}}
\newcommand{\douprule}{\rule[-2ex]{0pt}{4.5ex}}
\newcommand{\dorule}{\rule[-2ex]{0pt}{2ex}}
\def\thefootnote{\fnsymbol{footnote}}
\hyphenation{OGLE}
\begin{Titlepage}
\Title{The Optical Gravitational Lensing Experiment.\\
Gaia South Ecliptic Pole Field as Seen by \mbox{OGLE-IV}\footnote{Based on
observations obtained with the 1.3-m Warsaw telescope at the Las Campanas
Observatory of the Carnegie Institution for Science.}}
\vspace*{9pt}
\Author{I.~~S~o~s~z~y~ñ~s~k~i$^1$,~~
A.~~U~d~a~l~s~k~i$^1$,~~
R.~~P~o~l~e~s~k~i$^1$,~~
S.~~K~o~z~³~o~w~s~k~i$^1$,\\
£.~~W~y~r~z~y~k~o~w~s~k~i$^{1,2}$,~~
P.~~P~i~e~t~r~u~k~o~w~i~c~z$^1$,~~
M.\,K.~~S~z~y~m~a~ñ~s~k~i$^1$,\\
M.~~K~u~b~i~a~k$^1$,~~
G.~~P~i~e~t~r~z~y~ñ~s~k~i$^{1,3}$,~~
K.~~U~l~a~c~z~y~k$^1$~~
and~~
J.~~S~k~o~w~r~o~n$^{4,1}$}
{$^1$Warsaw University Observatory, Al.~Ujazdowskie~4, 00-478~Warszawa, Poland\\
e-mail: (soszynsk,udalski,rpoleski,simkoz,wyrzykow,pietruk,msz,mk,pietrzyn,\\
kulaczyk,jskowron)@astrouw.edu.pl\\
$^2$Institute of Astronomy, University of Cambridge, Madingley Road, Cambridge CB3~0HA,~UK\\
$^3$Universidad de Concepción, Departamento de Astronom\'ia, Casilla 160--C, Concepción, Chile\\
$^4$Department of Astronomy, Ohio State University, 140 W.~18th Ave., Columbus, OH~43210, USA}
\vspace*{5pt}
\Received{September 28, 2012}
\end{Titlepage}

\vspace*{5pt}
\Abstract{We present a comprehensive analysis of the Gaia South Ecliptic
Pole (GSEP) field, 5.3~square degrees area around the South Ecliptic Pole
on the outskirts of the LMC, based on the data collected during the fourth
phase of the Optical Gravitational Lensing Experiment, \mbox{OGLE-IV}. The GSEP
field will be observed during the commissioning phase of the ESA Gaia space
mission for testing and calibrating the Gaia instruments.

We provide the photometric maps of the GSEP region containing the mean {\it
VI} photometry of all detected stellar objects and their equatorial
coordinates. We show the quality and completeness of the \mbox{OGLE-IV} photometry
and color--magnitude diagrams of this region.

We conducted an extensive search for variable stars in the GSEP field
leading to the discovery of 6789 variable stars. In this sample we found
132 classical Cepheids, 686 RR~Lyr type stars, 2819 long-period, and 1377
eclipsing variables. Several objects deserving special attention were also
selected, including a new classical Cepheid in a binary eclipsing system.

To provide empirical data for the Gaia Science Alert system we also
conducted a search for optical transients. We discovered two firm type Ia
supernovae and nine additional supernova candidates. To facilitate future
Gaia supernovae detections we prepared a list of more than 1900 galaxies to
redshift about 0.1 located in the GSEP field.

Finally, we present the results of astrometric study of the GSEP field.
With the 26 months time base of the presented here \mbox{OGLE-IV} data,
proper motions of stars could be detected with the accuracy reaching
2~mas/yr. Astrometry allowed to distinguish galactic foreground variable
stars detected in the GSEP field from LMC objects and to discover about 50
high proper motion stars (proper motion $\gtrsim100$~mas/yr). Among them
three new nearby white dwarfs were found.

All data presented in this paper are available to the astronomical
community from the OGLE Internet archive.}{Techniques: photometric --
Astrometry -- Catalogs -- Stars: variables: general -- Stars: oscillations
(including pulsations) -- binaries: eclipsing -- Magellanic Clouds}

\vspace*{-7pt}
\Section{Introduction}
\vspace*{-5pt}
The Gaia satellite mission is a flagship scientific mission of the European
Space Agency to be launched in the second half of 2013. This is a very
ambitious project with the main scientific goal to provide very precise
astrometry (accuracy up to 20~$\mu$as for 15~mag stars), broad band
photometry and low resolution spectrophotometry covering wavelength range
300--1000 nm of about 1 billion stars in the Galaxy and Local Group as well
as radial velocity measurements of brighter objects from this sample (de
Bruijne 2012).

The Gaia hardware will consist of two telescopes with an aperture of
$1.45\times0.5$~m each imaging the sky on the focal plane where the main
scientific instrument consisting of 106 thin CCDs, in total almost 1
billion pixels, is located. The focal plane CCDs will be divided to sky
mapping, astrometric, blue and red spectrophotometric and radial
velocity arrays. The satellite is supposed to work in continuous sky
scanning mode, so the images will drift continuously through the
subsequent arrays filling the entire focal plane where the
appropriate measurements will be carried out. The satellite will be
placed on the orbit around the Sun-Earth L2 Lagrange point, about 1.5
million km from the Earth. The mission duration is supposed to be at
least five years.

After launching, during the commissioning phase, the Gaia satellite will be
operated in the scanning mode different from the nominal one to observe
more frequently some regions of the sky for calibration purposes and
testing the performance of instrumentation. It will cover the ecliptic
poles collecting data with higher than typical cadence. One of these
regions, namely located close to the southern ecliptic pole, falls on the
outskirts of the Large Magellanic Cloud -- the sky area monitored regularly
by the Optical Gravitational Lensing Experiment in its fourth
phase \mbox{OGLE-IV}. Throughout this paper we will call this region of the
LMC as the Gaia South Ecliptic Pole field (GSEP).

The GSEP field consists of four \mbox{OGLE-IV} pointings and has been
monitored regularly in the standard {\it VI} bands since 2010. Since then
the number of epochs collected in the {\it I}-band reached several
hundreds. Several tens of {\it V}-band observations were also secured for
color information. This dataset is large enough to perform complete
analysis of the field as seen from the ground and to provide its full
characterization. For example, it is possible to conduct very extensive and
effective search for variable stars, optical transients and analysis of
stellar populations. The long time span of observations also allows the
determination of precise astrometry of the field. In all these applications
the Gaia satellite is supposed to bring new and more precise space
measurements, so the direct comparison with the most precise ground base
data will be crucial for evaluation of the actual performance of the
satellite.

The OGLE photometry is regarded as one of the most precise available from
the ground. Moreover, the first results from the \mbox{OGLE-IV} phase indicate
that it outperforms that collected in the previous phases. The \mbox{OGLE-IV}
photometry of the GSEP field may be, then, an ideal observational material
for the Gaia mission for comparison and tests when launched. Therefore we
decided to analyze and make the data of the GSEP region of the LMC
available well ahead the entire \mbox{OGLE-IV} LMC data are released.

In the past, part of the GSEP region of the LMC was also observed from the
ground by the EROS microlensing survey (Tisserand \etal 2007) in the years
1996--2003. Additionally, the VISTA VMC Survey (Cioni \etal 2011) covered
the GSEP field in the near-infrared. However, these data can only be
complementary to the optical photometry by providing infrared colors. The
number of collected epochs is inadequate for more advanced applications as,
for example, a search for variable or transient objects.

Here we present a comprehensive analysis of the \mbox{OGLE-IV} data of the GSEP
region in the LMC. First, we show the quality of data, completeness of the
\mbox{OGLE-IV} photometry of this region of the LMC and color--magnitude 
diagrams of the GSEP sub-fields. Then, we present the results of an
extensive search for variable stars conducted in the GSEP field detecting
6789 variable objects down to $I\approx21$~mag. Among them 132 are
classified as classical Cepheids, 686 as RR~Lyr type stars, 2819 as
long-period and 1377 as eclipsing variables. We also searched for transient
objects detecting two firm supernovae and nine supernova candidates in the
background galaxies. For facilitating future detections of supernovae we
prepared a list of galaxies seen in the GSEP field. Finally, we present
proper motions of stars from the GSEP field which with the time baseline of
the \mbox{OGLE-IV} observations reach the precision better than
2~mas/yr. Astrometry allowed to distinguish foreground galactic objects
from genuine LMC stars. Also a few high proper motion stars and nearby
foreground white dwarfs were found. The accuracy of astrometry should
improve with time and at the time of the Gaia operation should reach
1~mas/yr.

All the \mbox{OGLE-IV} data presented here are available to the astronomical
community from the OGLE Internet archive.
\begin{figure}[t]
\centerline{\includegraphics[width=12.3cm]{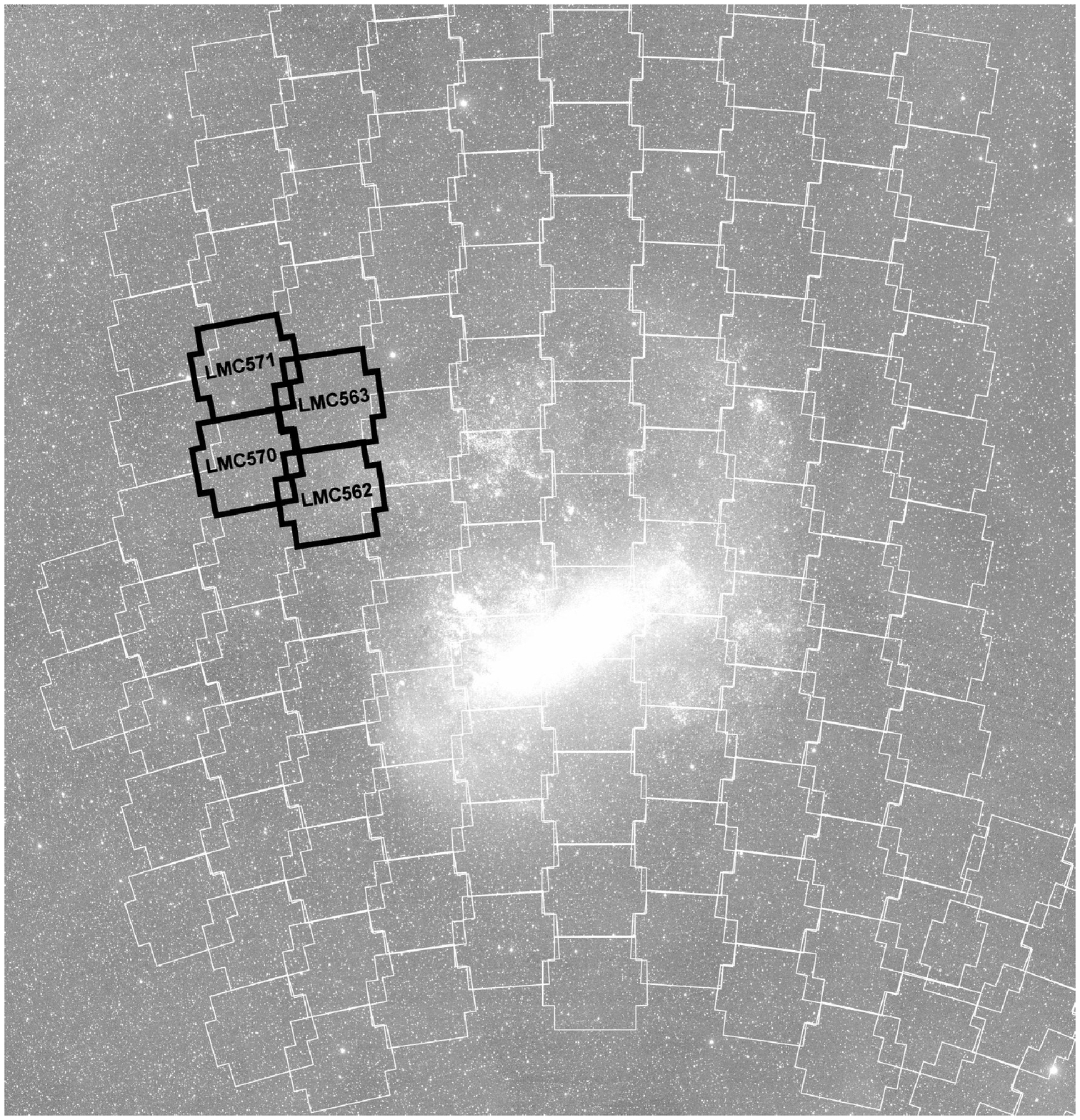}}
\vskip4mm
\FigCap{Gaia South Ecliptic Pole field in the LMC (black contours).
White contours show other \mbox{OGLE-IV} fields in the LMC. The background image
of the LMC was taken by the ASAS wide field sky survey (Pojmañski 1997).}
\end{figure}

\vspace*{-7pt}
\Section{Observational Data}
{\vspace*{-5pt}
The \mbox{OGLE-IV} survey is conducted with the 1.3-m Warsaw telescope at the Las
Campanas Observatory in Chile. The observatory is operated by the Carnegie
Institution for Science. During the \mbox{OGLE-IV} phase the Warsaw telescope is
equipped with the ``third generation'' mosaic camera with 32 thin E2V44-82
$2048\times4102$ pixel CCD detectors. The new camera covers approximately
1.4 square degrees on the sky with the scale of 0.26~arcsec/pixel. Detailed
description of the \mbox{OGLE-IV} instrumentation can be found at the OGLE WWW
site:

\vspace*{3pt}
\centerline{\it http://ogle.astrouw.edu.pl/main/OGLEIV/mosaic.html}
\vspace*{3pt}

The GSEP field consists of four \mbox{OGLE-IV} pointings with the standard
\mbox{OGLE-IV} designations: LMC562, LMC563, LMC570 and LMC571. Due to high
southern declination of the LMC fields and the constraints resulting from
the sky mosaicing pattern and the shape of the camera field of view, the
individual fields overlap quite significantly in some parts. Fig.~1
presents the location of the GSEP field in the LMC. Table~1 lists the
equatorial coordinates of the centers of the four GSEP subfields. They
cover an area of about 5.3 square degrees. The total number of stars
detected in the {\it I}-band exceeds 1.6~million -- the appropriate numbers
for each subfield are also listed in Table~1.

\renewcommand{\arraystretch}{0.9}
\MakeTableee{ccccrr}{12.5cm}{\mbox{OGLE-IV} pointings in the GSEP Field}
{\hline
\noalign{\vskip3pt}
Field & RA       &   DEC   & $N_{\rm Stars}$ \\
    & (2000)   &  (2000) & \\
\noalign{\vskip3pt}
\hline
\noalign{\vskip3pt}
LMC562  & 5\uph55\upm28\ups & $-67\arcd27\arcm45\arcs$ & 605\,304 \\
LMC563  & 5\uph53\upm47\ups & $-66\arcd13\arcm55\arcs$ & 445\,912 \\
LMC570  & 6\uph05\upm56\ups & $-66\arcd50\arcm50\arcs$ & 342\,033 \\
LMC571  & 6\uph03\upm51\ups & $-65\arcd37\arcm00\arcs$ & 272\,300 \\
\noalign{\vskip1pt}
\hline}
\renewcommand{\arraystretch}{1}

\vspace*{-7mm}
The GSEP subfields were photometrically monitored from March 6, 2010. The
last observations used in this analysis were collected on June 6,
2012. During that time we secured between 338 and 351 data points in the
Cousins {\it I}-band and 29 points in the {\it V}-band. The exposure time
was 150 seconds for both {\it V}- and {\it I}-band observations.

Photometry was obtained using the \mbox{OGLE-IV} photometric pipeline designed for
real time data reduction at the telescope. It is based on the \mbox{OGLE-III} data
pipeline (Udalski 2003) and the photometry is derived with the Difference
Image Analysis (DIA) technique (Alard and Lupton 1998, Wo¼niak 2000).

The {\it I}-band magnitudes of stars measured by OGLE range from 13~mag to
21~mag. The basic astrometric transformations between the pixel grid and
equatorial coordinates were based on the 2MASS Catalog coordinate grid as
in the case of the \mbox{OGLE-III} photometric maps (Szymañski \etal 2011).

\Section{\mbox{OGLE-IV} Photometry of the Gaia South Ecliptic Pole Field}
\vspace*{-5pt}
The GSEP field is located on the outskirts of the LMC and belongs to
relatively empty fields. Photometry of stellar objects detected on the {\it
I}-band reference images was stored in the individual databases for each of
the \mbox{OGLE-IV} GSEP subfields and CCD detectors (\eg lmc562.01, ...,
lmc562.32). Instrumental photometry was calibrated to the standard {\it VI}
system in two steps. First, we used the \mbox{OGLE-III} LMC Photometric
Maps (Udalski \etal 2008) as a huge list of secondary standards. Comparison
of the \mbox{OGLE-IV} photometry of the fields fully overlapping with
the \mbox{OGLE-III} maps collected on several photometric nights allowed to
derive color-term of \mbox{OGLE-IV} filters, large scale photometry
correction resulting among other factors from the change of the scale in
the large field of view of the OGLE camera, and the photometry zero
point. Only bright and non-variable stars were selected for this
comparison. This step required, of course, prior conversion of
the \mbox{OGLE-IV} database photometry to the aperture photometry scale by
the determination of appropriate aperture corrections.

\begin{figure}[p]
\vglue-5mm
\centerline{\includegraphics[width=9.5cm, bb=30 50 510 550]{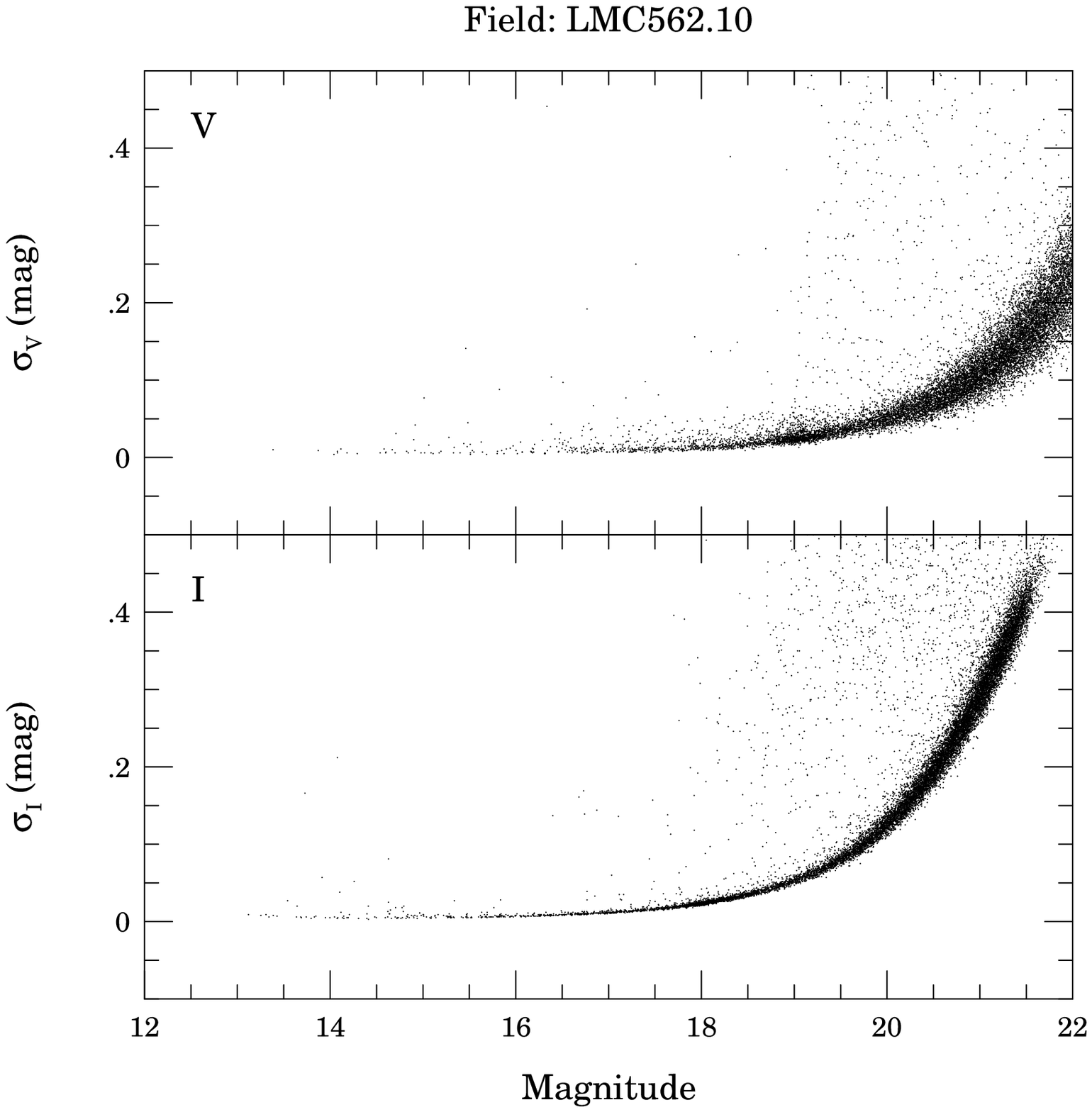}}
\centerline{\includegraphics[width=9.5cm, bb=30 50 510 550]{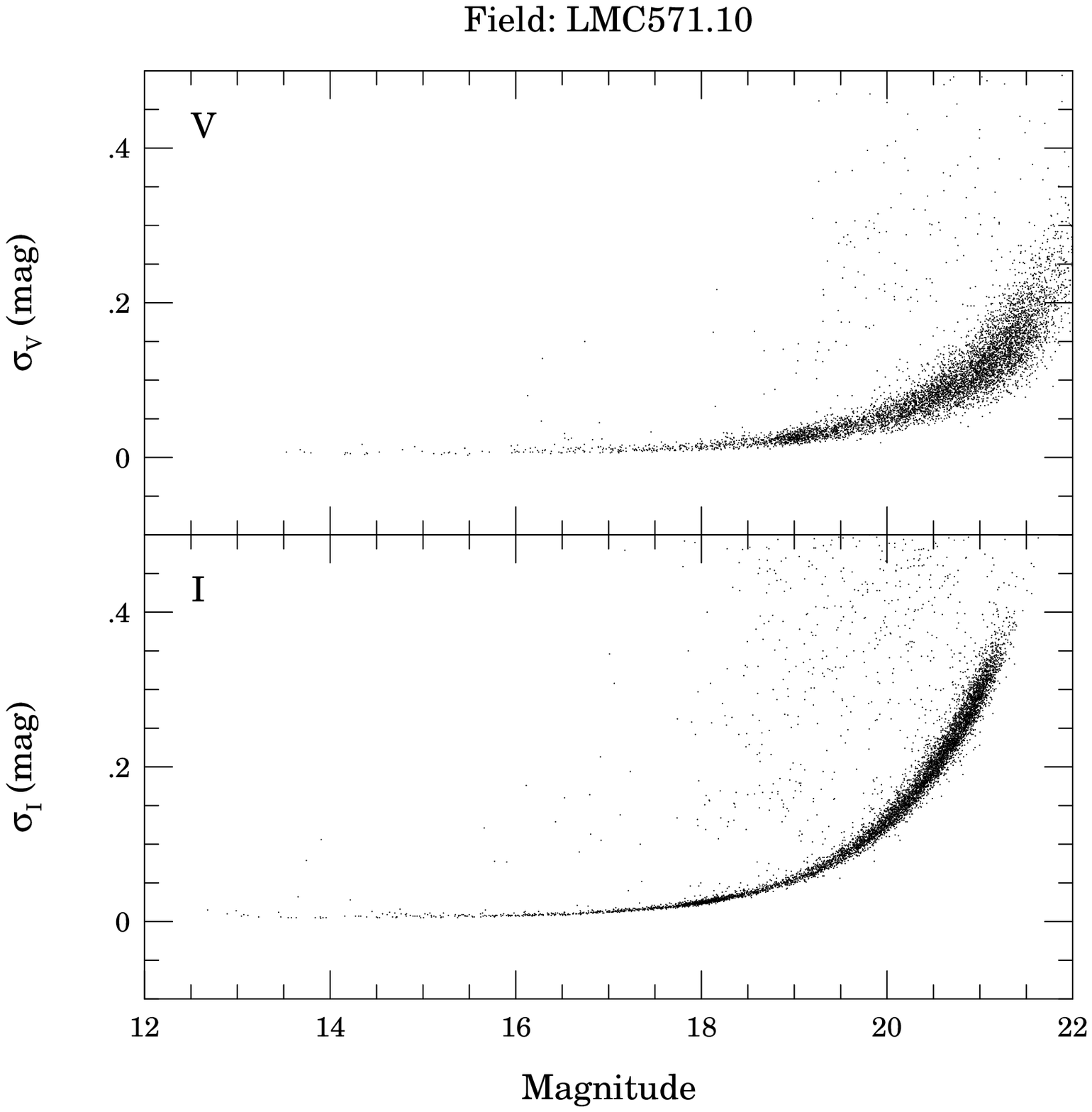}}
\FigCap{Standard deviation of magnitudes as a function of magnitude for
the subfields LMC562.10 (denser region in the GSEP field) and LMC571.10
(more sparse region).}
\end{figure}
\begin{figure}[p]
\vglue-5mm
\centerline{\includegraphics[width=9.5cm, bb=30 50 510 550]{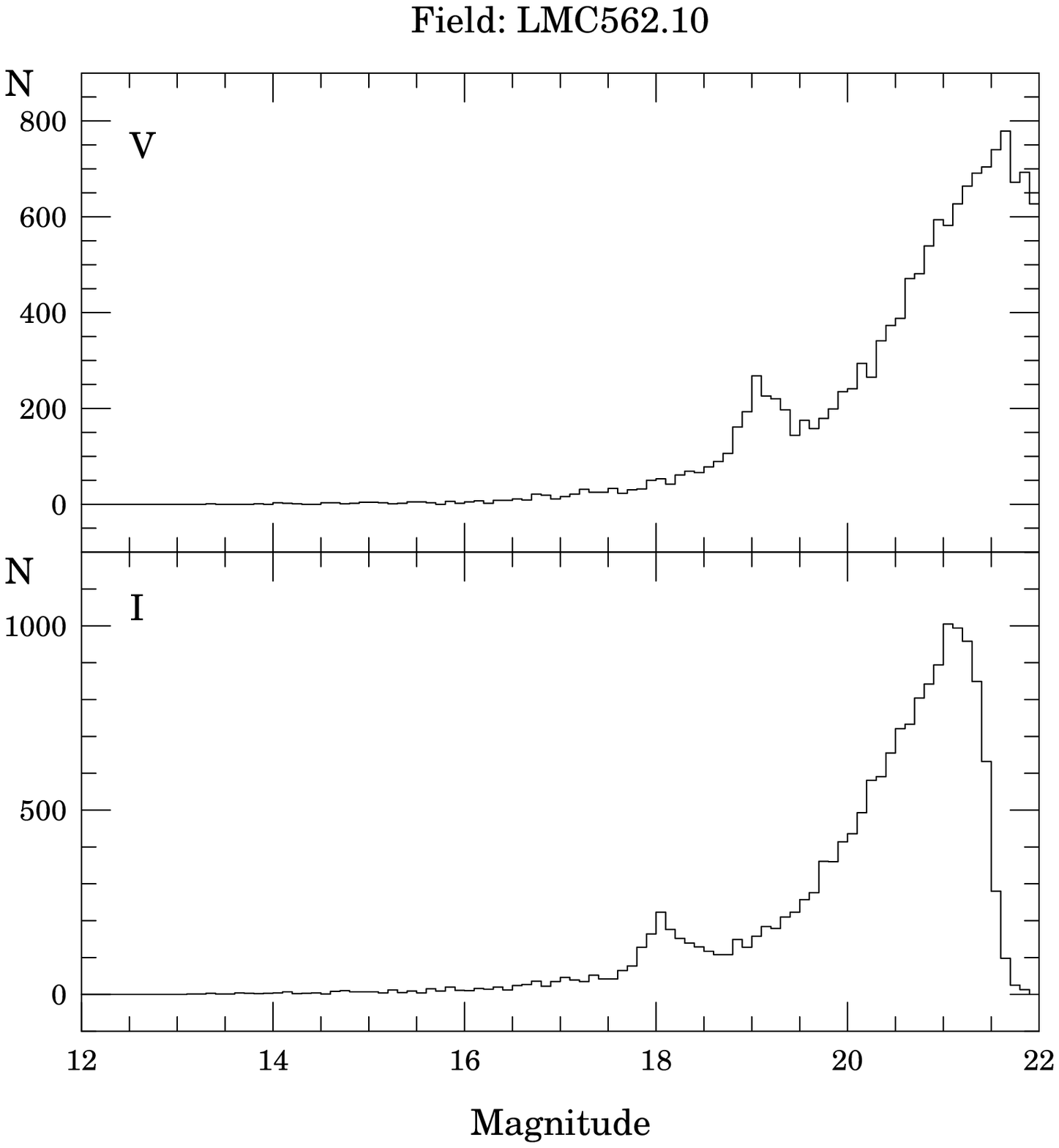}}
\centerline{\includegraphics[width=9.5cm, bb=30 50 510 550]{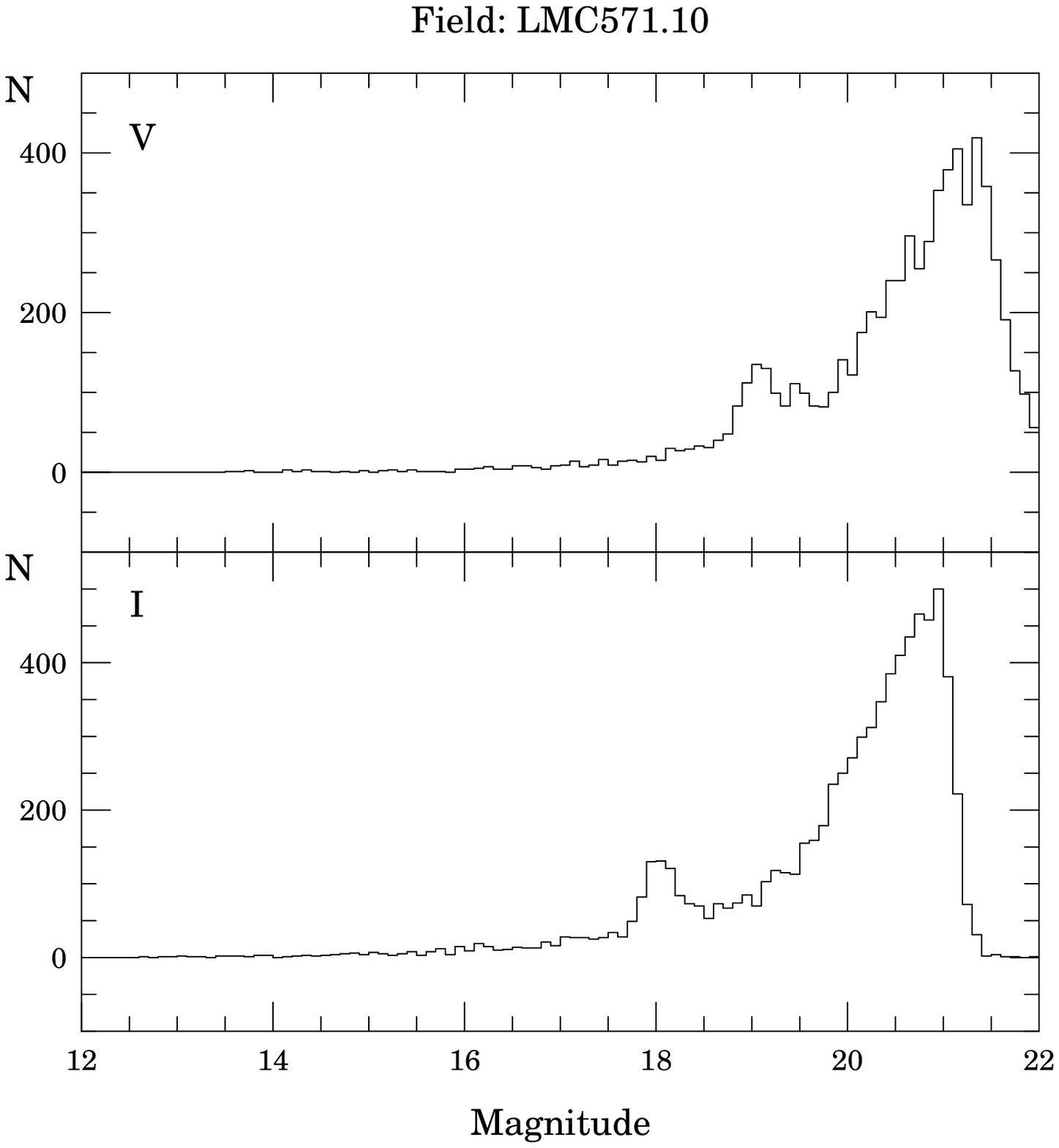}}
\FigCap{Histograms of star counts as a function of magnitude in the
subfields LMC562.10 and LMC571.10.}
\end{figure}
\begin{landscape}
\begin{figure}[p]
\vglue-10pt
\begin{tabular}{ll}
\includegraphics[height=12.3cm, bb=10 50 560 750]{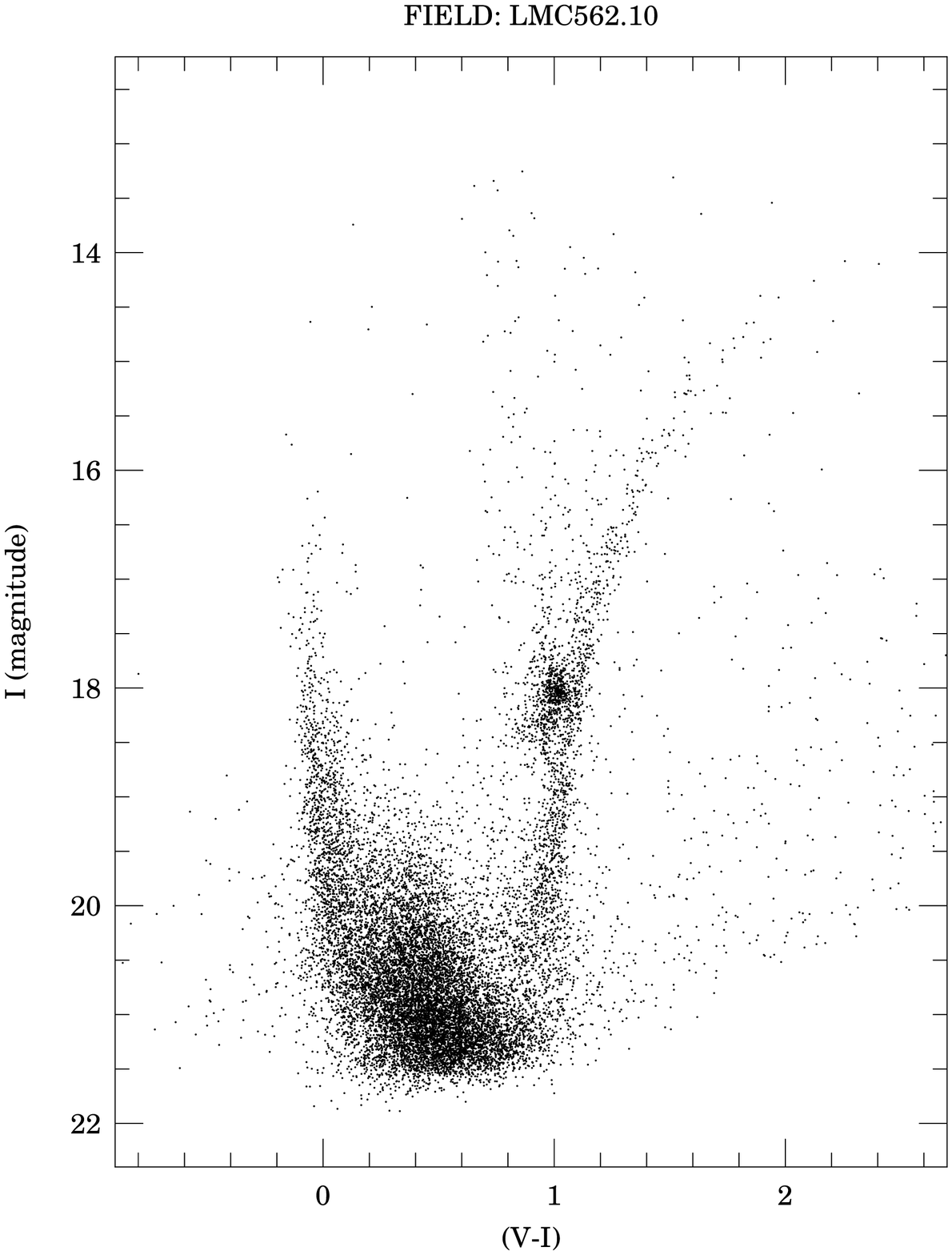}
&
\includegraphics[height=12.3cm, bb=10 50 560 750]{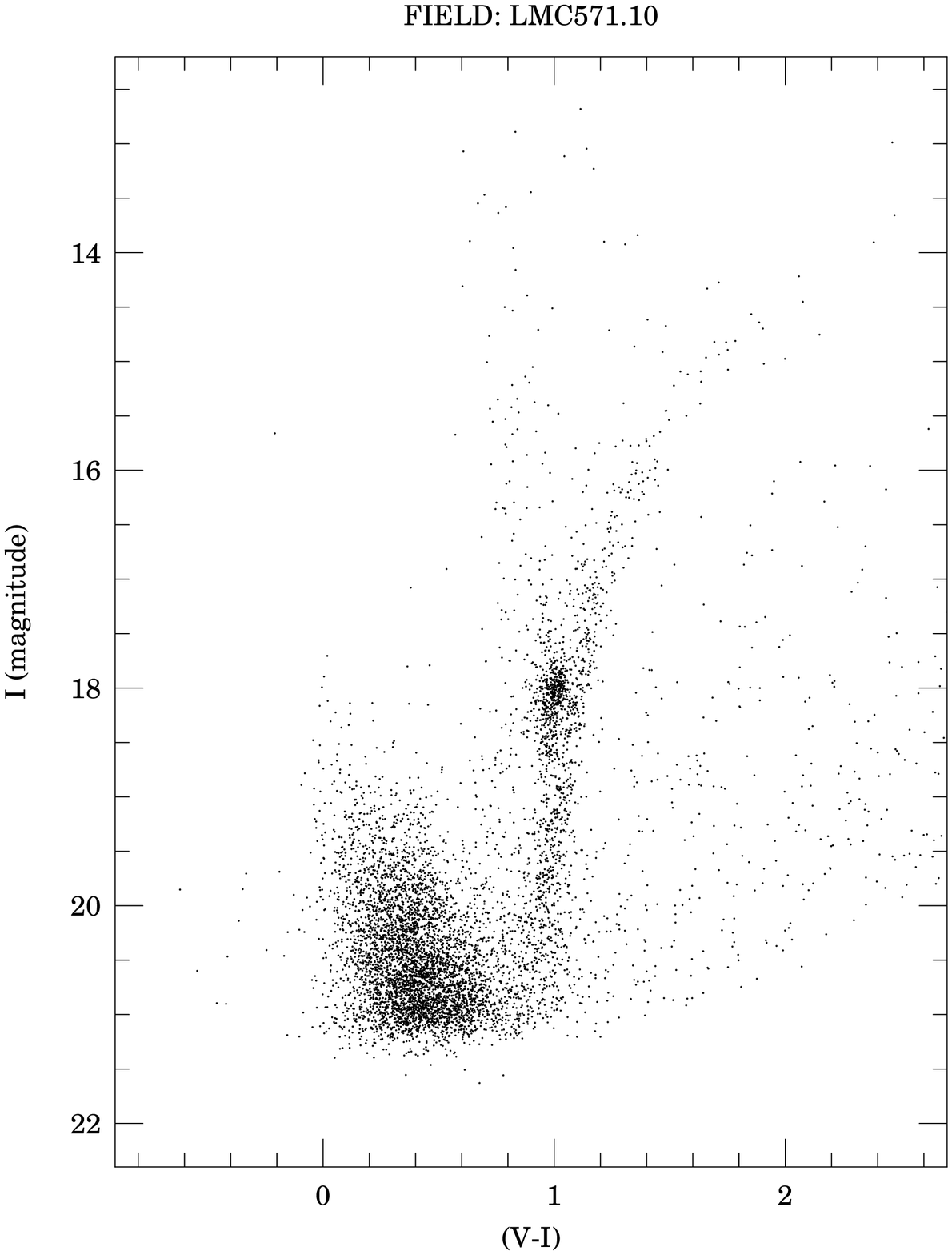}
\\
\FigCap{Color--magnitude diagram of the subfield LMC562.10.}
&
\hglue5mm{\FigCap{Color--magnitude diagram of the subfield LMC571.10.}}
\\
\end{tabular}
\end{figure}
\end{landscape}

Having the transformation parameters from the comparison of the \mbox{OGLE-IV}
photometry with photometric standards (\mbox{OGLE-III} maps) derived in the
previous step, the appropriate transformations to the photometry of the
GSEP images were applied for each photometric night. The final calibration
of the instrumental GSEP databases was the average of individual
calibrations from these nights. We estimate the accuracy of zero points of
the final photometry to be about 0.03~mag.

Average {\it VI} photometry of all objects detected in the {\it I}-band
reference images combined with the derived equatorial coordinates, namely
the \mbox{OGLE-IV} photometric maps of the GSEP field, were prepared in the
same manner as for the \mbox{OGLE-III} maps (Szymañski \etal 2011). The
maps provide information for fast assessment of the quality of
the \mbox{OGLE-IV} photometry. Fig.~2 shows the standard deviation as a
function of magnitude for typical subfields LMC562.10 (denser region of the
GSEP field) and LMC571.10 (sparse stellar density) for {\it V}- and {\it
I}-band. As one can see the \mbox{OGLE-IV} accuracy of photometry falls to
0.1~mag at $V\approx 21$~mag and $I\approx20$~mag which is roughly
comparable with the Gaia expected performance.

Fig.~3 presents histograms of the number of detected stars as a function of
magnitude in 0.1~mag bins for the same subfields. The \mbox{OGLE-IV} photometry is
complete to about 21.5~mag and 21.0~mag in the {\it V}- and {\it I}-band,
respectively.

Figs.~4 and 5 present color--magnitude diagrams of the LMC562.10
and\linebreak LMC571.10 subfields. Both show typical features of the CMDs
from the LMC fields (\cf Udalski \etal 2008) including the main sequence
stars strip, red giant branch, and prominent red clump.

Large number of good resolution images collected during the \mbox{OGLE-IV}
survey allows construction of much deeper color--magnitude diagrams by
stacking such high quality images. Although not done yet, it is planned to
make deep OGLE maps of the all observed fields in the Magellanic Clouds.

\Section{Search for Variable Stars}
Search for variable objects in the GSEP field began with the period search
conducted on the {\it I}-band light curves of all detected stars. We used
the {\sc Fnpeaks} program (Z.\,Ko³aczkowski, private communication) which
analyzes the Fourier spectra of the light curves. The search was performed
in the frequency range 0~--~24~day$^{-1}$, with the resolution of
$5\times10^{-5}$~day$^{-1}$. To avoid daily aliases the period search for
long-period variables was limited to the maximum frequency of
0.5~day$^{-1}$.

The selection and classification of variable stars was based primarily on
the light curve shapes. We visually inspected all {\it I}-band light curves
brighter than $I=17$~mag. For fainter stars we examined the light curves
with signal to noise of the most prominent peak in the periodogram larger
than 4.5. Our final classification took into account the morphology of the
light curves, mean {\it I}-band magnitudes, $(V-I)$ colors, near-infrared
{\it JHK} magnitudes from the 2MASS Point-Source Catalog (Cutri
\etal 2003) and ratios of periods (for multi-periodic stars). In total we
visually examined 135\,172 light curves, finding 6789 variable stars.

\begin{figure}[p]
\hglue-11mm{\includegraphics[width=14.6cm]{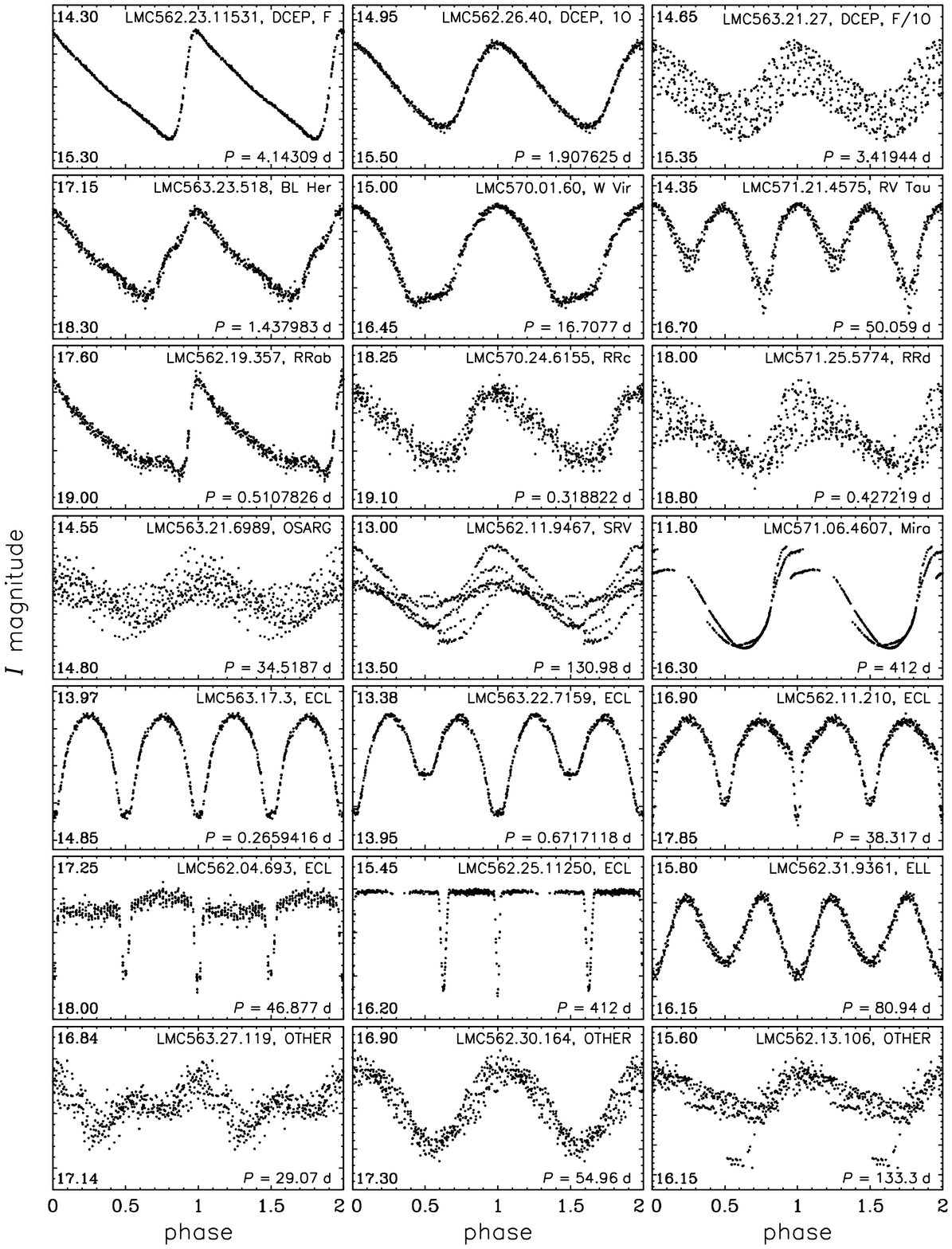}}
\vspace*{-14mm}
\FigCap{Examples of the {\it I}-band light curves of periodic variable
stars from the GSEP field. Successive rows show the following types of
variability: classical Cepheids, type~II Cepheids, RR~Lyr stars,
long-period variables, eclipsing and ellipsoidal binary systems and other
variables. Note that the range of magnitudes varies from panel to panel.
Numbers in the left corners show the minimum and maximum magnitudes of the
range.}
\end{figure}

\Subsection{Catalog of Variable Stars in the GSEP Field}
All stars detected in the GSEP field of the LMC and classified as variables
are listed in the Catalog of Variable Stars in the GSEP Field which is
described below. It is supposed to be a part of the main OGLE Catalog of
Variable Stars which will be published in the future, when \mbox{OGLE-IV}
data will be gradually released. It is provided in the electronic form in
the OGLE Internet archive (see Section~7) and as such it is expected to be
updated when more data are available, more precise calibrations are
derived, and some unavoidable errors are noticed and corrected.

The list of all variable stars is given in the file {\sf
list.dat}. This file contains a table with the following columns:
star ID (LMCNNN.MM.KKK, where NNN is the field number, MM indicates the CCD
chip number of the mosaic camera, and KKK is the star number in the
\mbox{OGLE-IV} database), J2000 equatorial coordinates, type of
variability, subtype or secondary type of variability, mean {\it I}- and
{\it V}-band magnitudes, period for periodic stars -- derived with the {\sc
Tatry} code kindly provided by Schwarzenberg-Czerny (1996), peak-to-peak
{\it I}-band amplitude, and the secondary period for double- and
multi-periodic stars. Cross-identifications with the General Catalogue of
Variable Stars (GCVS) and remarks on interesting objects are given in the
file {\sf remarks.dat}.

The time-series {\it I}- and {\it V}-band photometry of the stars is stored
in the directory {\sf phot/}. Finding charts for all stars can be
downloaded from the directory {\sf fcharts/}. These are
$60\arcs\times60\arcs$ subframes of the {\it I}-band DIA reference
images. See the {\sf README} file for more details. Figs.~6 and~7 present
examples of the periodic and non-periodic light curves of variable stars
from the GSEP field catalog.
\begin{figure}[t]
\vglue-0.8cm
\centerline{\includegraphics[width=13.8cm]{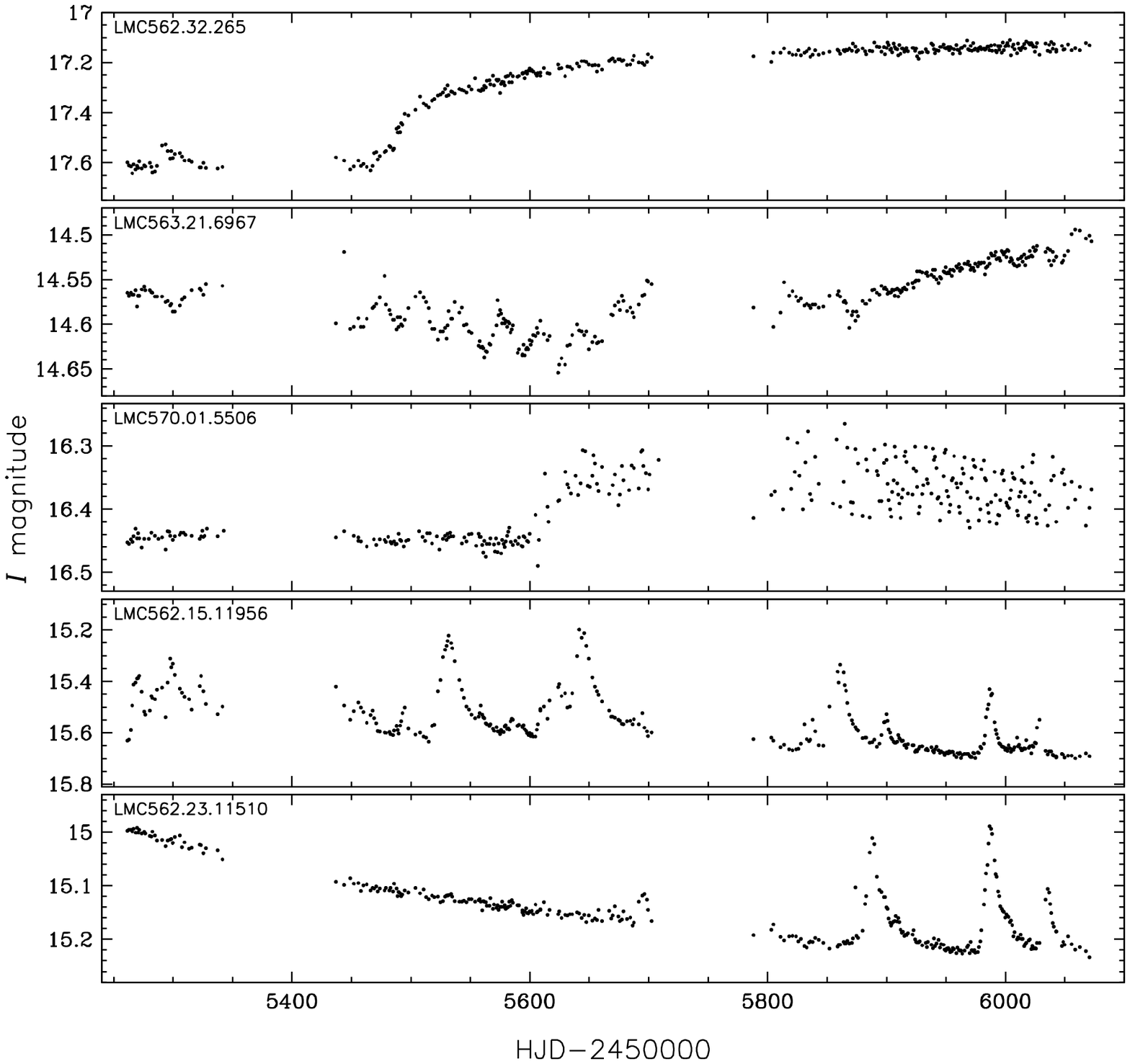}}
\vspace*{-53mm}
\FigCap{Examples of the {\it I}-band light curves of non-periodic variables
from the Catalog of Variable Stars in the GSEP Field.}
\end{figure}

\Subsection{Completeness of the Catalog}
The completeness of our catalog of the GSEP field was assessed by checking
our efficiency of the variable star selection in the overlapping parts of
adjacent fields. Assuming that the minimum number of observing points must
be larger than 50, in total 378 variable stars from our catalog were
recorded in the \mbox{OGLE-IV} databases twice -- in the neighboring
fields, so we had an opportunity to independently detect 756
counterparts. As a result of our variability search we found 646 of them,
which yields the completeness of our search method of 83\%. The
completeness of the whole catalog depends additionally on the efficiency of
the star detection in the OGLE fields, which strongly depends on their
brightness. The completeness for stars fainter than $I=21$~mag drops
rapidly (see Section~3).

The verification of the missing 110 counterparts showed that they are
almost exclusively very faint or very low-amplitude stars, close to the
detection limits of the \mbox{OGLE-IV} survey. Light curves of the faintest
stars (with $I>20$~mag) are affected by large scatter, which lowers the
signal-to-noise ratios of their periodicities below the detection
thresholds applied in this study. The very low-amplitude stars were usually
missed during the visual examination. OGLE small amplitude red giants,
faint eclipsing binaries and $\delta$~Sct stars dominate in the group of
missed objects. We found all counterparts for Cepheids, RR~Lyr stars,
semiregular variables and Miras.

We also searched the GCVS Vol.V. Extragalactic Variable Stars (Artyukhina
\etal 1995) for previously known variable stars in the region covered by
our catalog. A total number of 70 variables was found in the GCVS, of which
48 objects were independently identified in our analysis. A careful
inspection of the missed 22 stars showed that: eight objects are too bright
and saturate in the OGLE frames, seven stars fell into the gaps between the
CCD chips of the \mbox{OGLE-IV} mosaic camera, for five objects we cannot
confirm their variability (OGLE photometry shows non-variable stars within
errors), and two objects exhibit low-amplitude variations, close to our
detection limits. These two stars were added to the present catalog and
classified as OTHER variables. The comparison to the GCVS confirms that the
completeness of our catalog is high, with the exception of faint or
low-amplitude stars.

\Subsection{\mbox{OGLE-IV} Variable Stars in the GSEP Field}
Table~2 lists the numbers of variable stars of different types included in
the catalog. In some cases our classification is not unique, since some
stars exhibit two types of variations, usually originated from pulsations
and binarity. For example our catalog contains a classical Cepheid showing
eclipses or ellipsoidal variables exhibiting OGLE Small Amplitude Red Giant
(OSARG) oscillations.
\MakeTableee{lcc}{12.5cm}{Number of variable stars of different variability types}
{\hline
\noalign{\vskip3pt}
Variability type & Flag & Number \\
    &    & of stars \\
\noalign{\vskip3pt}
\hline
\noalign{\vskip3pt}
classical Cepheids    & DCEP  & 132 \\
type II Cepheids      & T2CEP & 5 \\
anomalous Cepheids    & ACEP  & 3 \\
RR~Lyr stars        & RRLYR & 686 \\
$\delta$~Sct stars  & DSCT  & 159 \\
long-period variables & LPV   & 2819 \\
eclipsing binaries    & ECL   & 1377 \\
ellipsoidal variables & ELL   & 156 \\
other variables       & OTHER & 1473 \\
\noalign{\vskip3pt}
\hline}

Below we present a short summary of the content of our GSEP Field
Catalog regarding different types of stellar variability.

\vspace*{5mm}
{\it Classical Cepheids}

\begin{figure}[t]
\vglue-0.9cm
\centerline{\includegraphics[width=13.4cm]{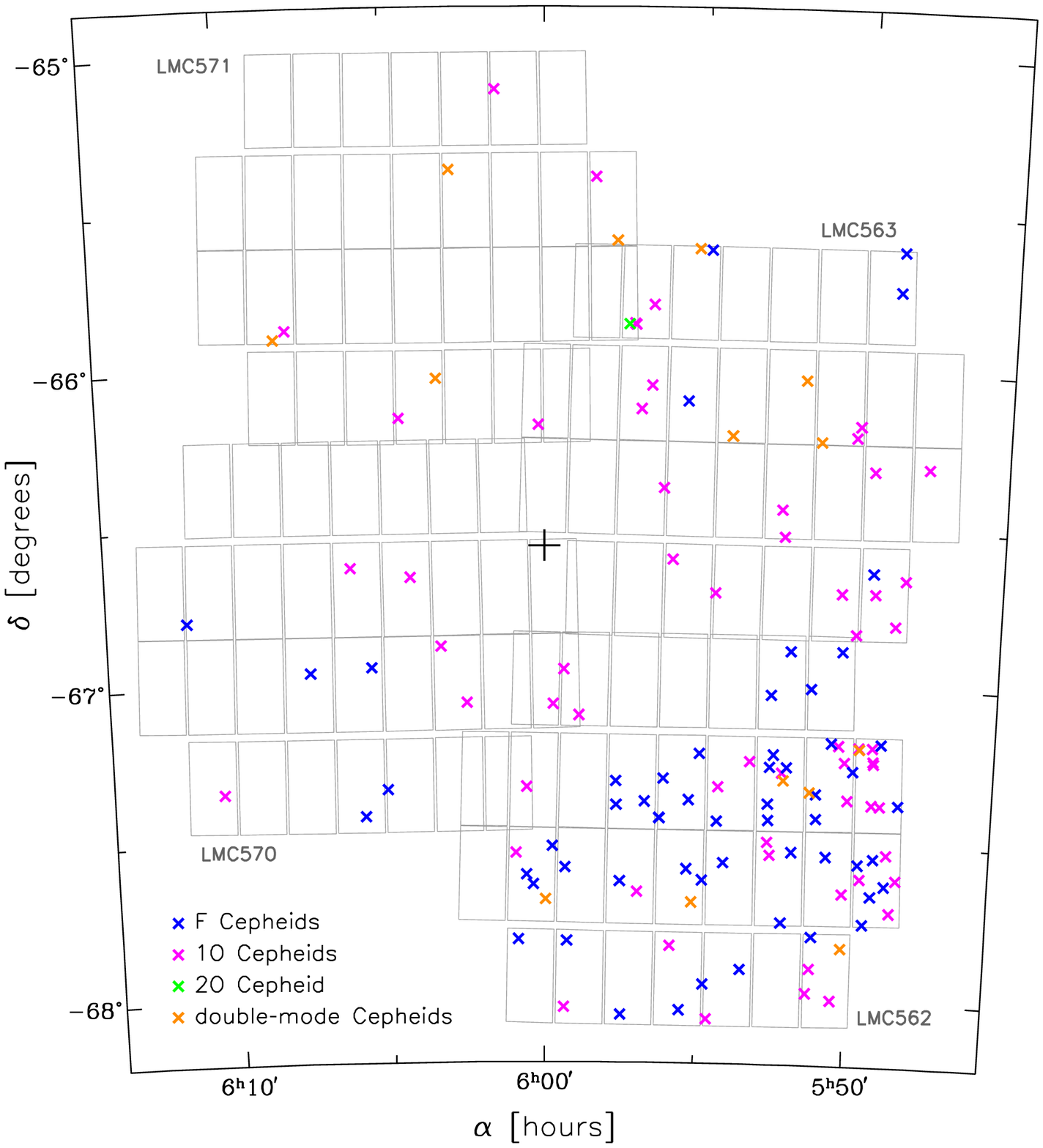}}
\vspace*{-35mm}
\FigCap{Spatial distribution of classical Cepheids in the GSEP field. Gray
contours show the sky coverage by the individual chips of the \mbox{OGLE-IV}
mosaic CCD camera. Blue, magenta, green and orange symbols represent
fundamental-mode, first-overtone, second-overtone and double-mode
pulsators, respectively. Black cross indicates the position of the South
Ecliptic Pole.}
\end{figure}
Most of the variable stars known to date in the region covered by this
study are classical Cepheids. The majority of them were discovered by the
Harvard survey for variable stars in the Magellanic Clouds (Leavitt 1908,
Shapley and Mohr 1933, Wetzel 1955). Our catalog contains 132 classical
Cepheids, 38 of them were included in the GCVS (five with wrong
classification). Fifty seven of our Cepheids pulsate solely in the
fundamental mode (F), sixty are single-mode first-overtone pulsators (1O),
one object is a single-mode second-overtone Cepheid (2O). Four stars are
double-mode F/1O Cepheids and ten objects pulsate simultaneously in the
first two overtones (1O/2O). Fig.~8 shows the spatial distribution of
classical Cepheids in the four \mbox{OGLE-IV} fields. Note the sharp edge
of the fundamental-mode Cepheid distribution crossing the LMC562
field. This reflects the distribution of the young stellar population in
the LMC, seen also in the CMDs (Figs.~4 and 5). The overtone and
double-mode Cepheids are distributed more homogeneously over the studied
area. The positions of double-mode Cepheids, RR~Lyr stars and $\delta$~Sct
stars in the Petersen diagram (\ie the plot of period-ratio against the
logarithm of the longer period) is shown in Fig.~9.
\begin{figure}[p]
\vglue-0.7cm
\centerline{\includegraphics[width=14.4cm]{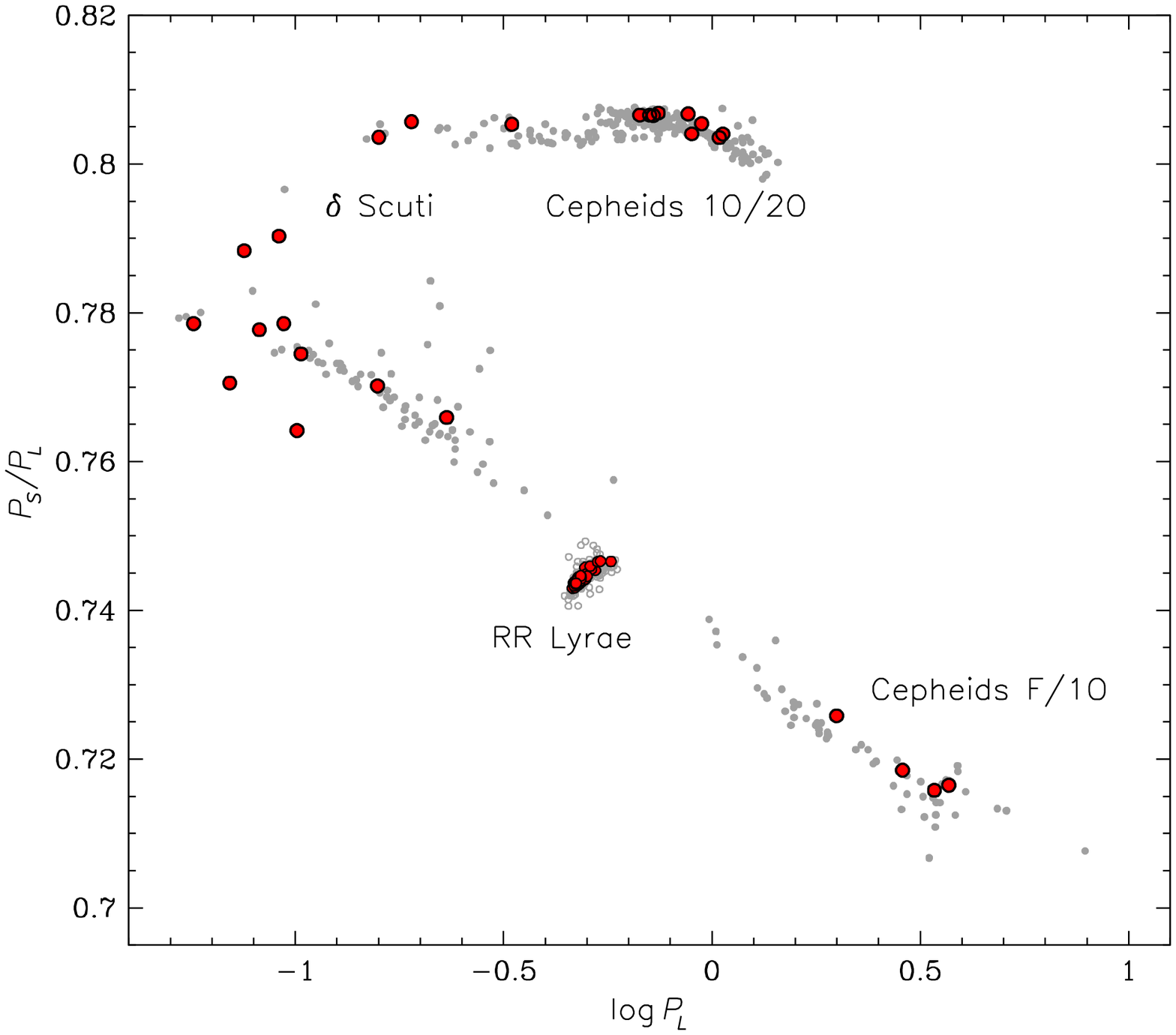}}
\vspace*{-60mm}
\FigCap{Petersen diagram for double-mode classical Cepheids, RR~Lyr stars
and $\delta$~Sct variables. Red symbols show objects in the GSEP field,
while gray points indicate LMC variables from the \mbox{OGLE-III} Catalog of
Variable Stars (Soszyñski \etal 2008, 2009a, Poleski \etal 2010a).}
\vskip5mm
\centerline{\includegraphics[width=14.3cm]{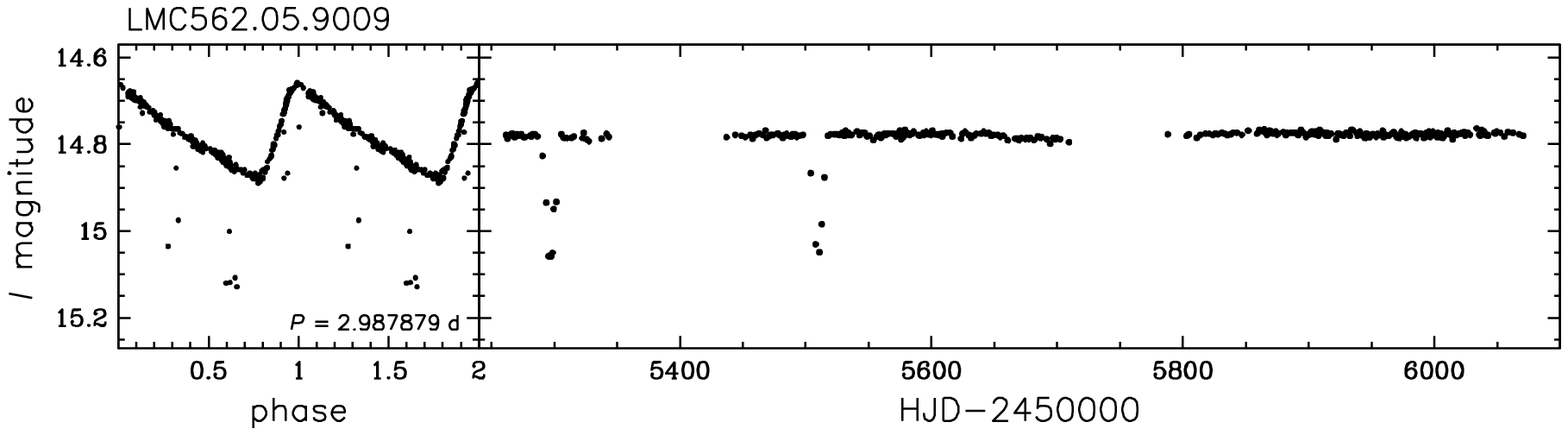}}
\vspace*{-143mm}
\FigCap{Light curve of a classical Cepheid LMC562.05.9009 exhibiting
eclipsing variations. {\it Left panel} shows the original photometric data
folded with the pulsation period. {\it Right panel} shows the unfolded
light curve after subtracting the Cepheid component.}
\end{figure}

We should stress here that an object designated as LMC562.05.9009 is worth
of particular interest. It belongs to a rare class of classical Cepheids in
eclipsing binary systems. Another Cepheid in the LMC (OGLE-LMC-CEP-0227)
which is a member of an eclipsing binary system was recently used for the
first determination of the dynamical Cepheid mass with the 1\% accuracy
(Pietrzyñski \etal 2010). OGLE-LMC-CEP-1812 system is another example of
such very rare systems (Pietrzyñski \etal 2011).

\hglue-7pt The light curve of LMC562.05.9009 is presented in Fig.~10. 
So far, the OGLE ob\-s\-er\-vations covered only two eclipses during the
seasons 2009/2010 and 2010/2011, and currently it is impossible to
determine the orbital period of the system. The lack of eclipses during the
2011/2012 season suggests an eccentric orbit. The system has already been
extensively followed-up both photometrically and spectroscopically.

\vspace*{5mm}
{\it Type II and Anomalous Cepheids}

The OGLE Catalog of the GSEP field contains five stars classified as
type~II Cepheids (one was already known). This sample consists of one
BL~Her star, one W~Vir star, two RV~Tau stars, and one yellow semiregular
variable (SRd). There are three anomalous Cepheids in our catalog (two
fundamental-mode and one first-overtone pulsator) and all of them are new
findings. The classification of LMC563.31.241 is uncertain, since it has a
short period (0.5575~days) -- typical for RR~Lyr stars. However, this
object is more luminous than a typical RR~Lyr star from the LMC, and its
light curve morphology resembles that of the fundamental-mode anomalous
Cepheids.

\begin{figure}[t]
\vglue-0.7cm
\centerline{\includegraphics[width=13.7cm]{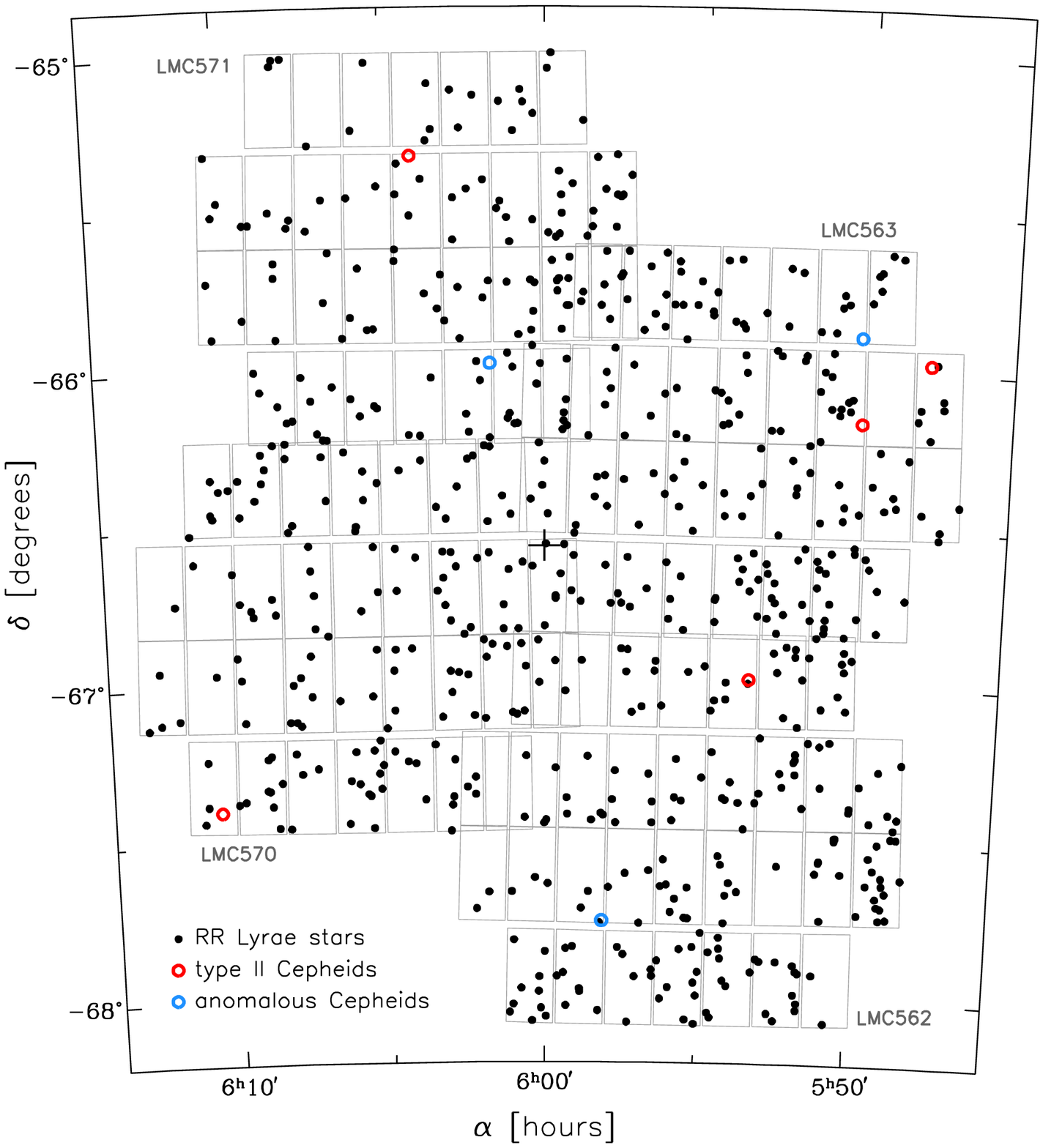}}
\vspace*{-35mm}
\FigCap{Spatial distribution of RR~Lyr stars, type II Cepheids and
anomalous Cepheids in the GSEP field. Gray contours show the sky coverage
by the individual chips of the \mbox{OGLE-IV} mosaic CCD camera. Black, red and
blue symbols show positions of RR~Lyr stars, type II Cepheids and anomalous
Cepheid, respectively. Black cross indicates the position of the South
Ecliptic Pole.}
\end{figure}

\vspace*{5mm}
{\it RR~Lyr and $\delta$~Sct Stars}

RR~Lyr stars are well known tracers of the old stellar population. The
\mbox{OGLE-III} catalog of RR~Lyr stars in the LMC (Soszyñski \etal 2009a)
contains the largest collection of these stars (24\,906) found in any
stellar environment. The present sample contains 686 RR~Lyr stars, nine of
which belong to the Milky Way and the remaining stars being members of the
LMC. Two Galactic RR~Lyr stars from this sample are present in the
GCVS. Our catalog contains 482, 164 and 40 RRab, RRc and RRd stars,
respectively. Hypothetical second-overtone pulsators (RRe stars) were not
distinguished from the RRc stars. The homogeneous spatial distribution of
RR~Lyr stars in our fields (Fig.~11) reflects the widely extended halo of
the old stellar population in the LMC.

The catalog of $\delta$~Sct stars in the LMC from the \mbox{OGLE-III}
project was published by Poleski \etal (2010a). Stars of this type usually
have magnitudes which are at the observing limits of the OGLE project
photometry, so for most of them our classification is rather uncertain. In
the GSEP region we detected 159 $\delta$~Sct stars. Among them 10 objects
show double-mode behavior, with the period ratios between 0.76 and 0.79 --
typical for fundamental-mode/first-overtone pulsators (Pigulski \etal
2006). Two $\delta$~Sct stars oscillate in the first and second
overtones. Fig.~9 shows newly found double-mode $\delta$~Sct stars in the
Petersen diagram, together with variables from the catalog of Poleski \etal
(2010a).

\begin{figure}[t]
\vglue-0.8cm
\centerline{\includegraphics[width=14cm]{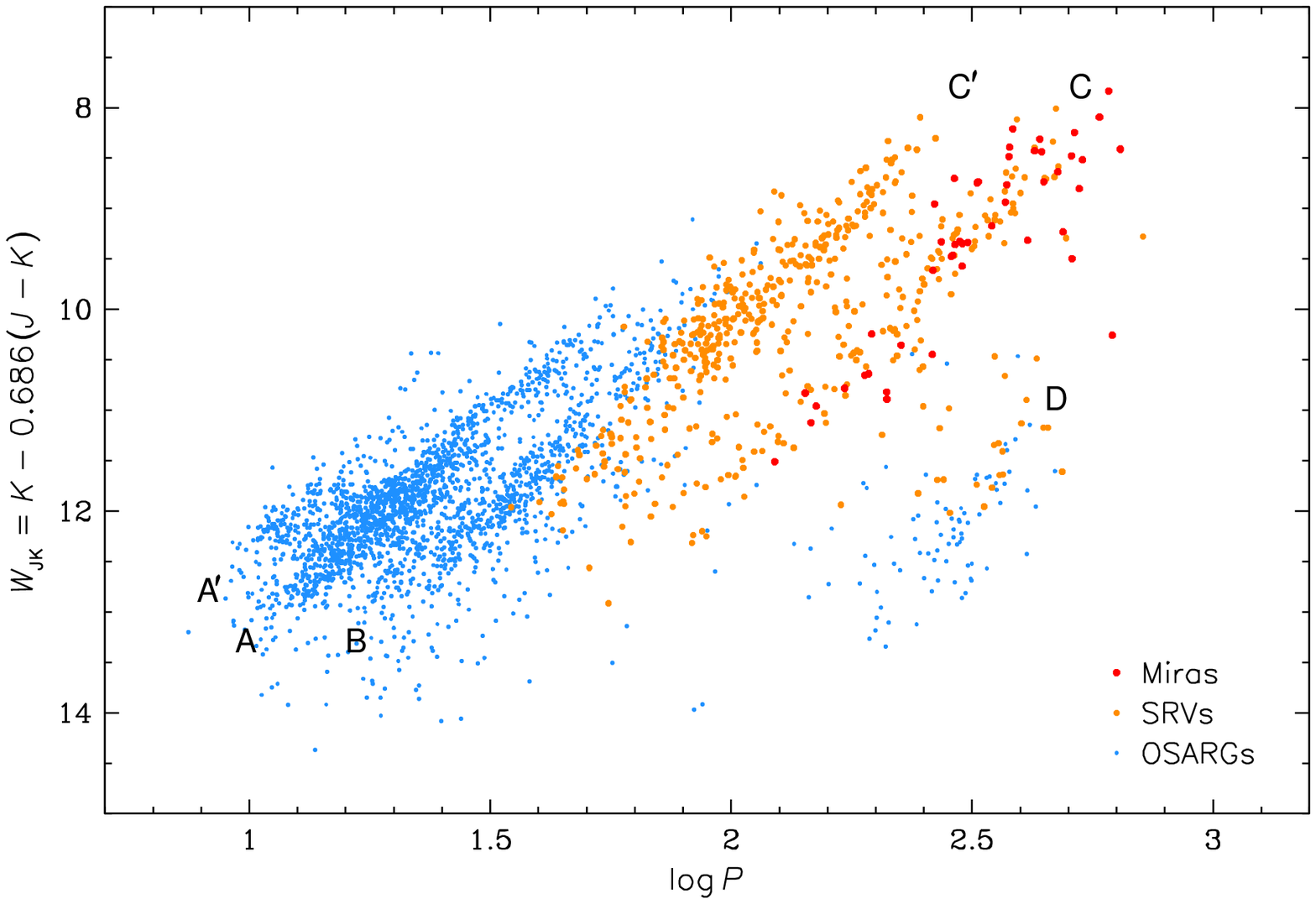}}
\vspace*{-88mm}
\FigCap{Near-infrared Wesenheit index \vs luminosity diagram for
long-period variables in the GSEP field. Blue, orange and red points
indicate OSARGs, SRVs and Mira stars. Labels show the PL sequences
introduced by Wood \etal (1999). Each star is represented by one, the
primary period. Near-infrared measurements were taken from the 2MASS
catalog (Cutri \etal 2003).}
\end{figure}

\vspace*{5mm}
{\it Long Period Variables}

Long-period variables (LPVs) constitute the most numerous group of variable
stars in the GSEP Field Catalog. In total we identified 2819 LPVs in the
four \mbox{OGLE-IV} fields. As in the \mbox{OGLE-III} catalog of LPVs in
the Magellanic Clouds (Soszyñski \etal 2009b), we divided our sample of
pulsating red giants into three groups: Miras, semiregular variables (SRVs)
and OGLE small amplitude red giants (OSARGs). We used the same criteria of
our classification, \ie as Miras we recognized stars with the peak-to-peak
amplitude of the detrended {\it I}-band light curve larger than 0.8~mag,
while OSARGs differ from SRVs by their separate period--luminosity
relations (in the {\it K}-band and Wesenheit indices) and characteristic
period ratios (Soszyñski \etal 2007). Note that in this work we applied
these criteria to the light curves spanning only two years, and in the
future, with more data and longer time span, our classification of some
individual objects may change.

The near-infrared period--luminosity diagram (as a ``luminosity'' we used
the reddening-free Wesenheit index, defined as $W_{JK}=K-0.686(J-K)$) for
LPVs (Fig.~12) reveals the well known pattern, noticed for the first time
by Wood \etal (1999). The only small difference can be seen for the
longest-period giants, in particular in sequence D populated by stars with
the so called long-secondary periods. This is due to relatively short time
baseline of our observations.

\vspace*{5mm}
{\it Eclipsing and Ellipsoidal Binaries}

Our search for variable stars resulted in the detection of 1533 binary
systems, usually eclipsing or ellipsoidal variables. All of them are new
findings. The Galactic star LMC570.22.45 shows the shortest orbital period
of 3.361 hours. The longest-period binaries are ellipsoidal red giants,
with periods reaching several hundred days. The distributions of colors and
magnitudes of our binary systems are similar to the distributions of the
whole population of stars in the OGLE database, which means that faint
objects dominate in the list of eclipsing binaries. Since the light curves
of faint stars are usually noisy, we have not divided our sample into
contact, semi-detached and detached systems.

Our catalog contains several interesting cases. LMC562.11.9469 shows
overlapping ellipsoidal and eclipsing light curves, with the orbital
periods of 421.5~days and 1.712147~days, respectively. At least ten
eclipsing binaries exhibit additional variations with periods close (but
not equal) to the orbital periods -- the feature typical for RS~CVn
variable stars. We flagged these objects in the catalog remarks. Four
eclipsing variable stars show additional periods from 29 to 40 times longer
than their orbital periods. These objects belong to the so called
double-periodic variables (DPVs) -- a class of binaries discovered by
Mennickent \etal (2003) in the OGLE data. Poleski (2010b) published the
catalog of 125 DPVs in the LMC.

\vspace*{5mm}
{\it Other Variable Stars}

Our catalog also contains 1473 periodic and non-periodic objects classified
as OTHER variable stars. Their variability type cannot be unambiguously
determined from the available data or the classification is uncertain. Many
of these objects show characteristic features of the rotating spotted
stars. This group likely contains also binary systems, Be stars,
$\delta$~Sct and other pulsating stars. Additional information like
spectral features could be conclusive in these cases for proper
interpretation.

\Section{\mbox{OGLE-IV} Support for Gaia Science Alerts System}
The data processing pipeline of Gaia is designed for near-real-time
alerting on detections of anomalies or brightenings or appearances of new
objects. The alerting system will be operating already during the
commissioning phase of Gaia, but will be thoroughly tested on these first
scientific Gaia data. The pipeline will exploit both photometry and
low-resolution spectroscopy to identify potential transient events
(Wyrzykowski and Hodgkin 2012).

As the \mbox{OGLE-IV} GSEP field can be the first testing point for the Gaia Alert
system, we took a closer look at the transients that can be detected and
expected in this region of the sky. We also prepared ground based data
facilitating the detection and interpretation of potential supernovae by
Gaia.

\Subsection{Search for \mbox{OGLE-IV} Transient Objects in the GSEP Field}
Apart from searching for ordinary variable stars, we conducted independent
search of the \mbox{OGLE-IV} databases for transient events. We searched both the
main \mbox{OGLE-IV} databases and the databases of new objects. The latter contain
objects that are not present on the DIA reference images, so they seem to
appear from below the detection threshold level. In such sparse stellar
density fields, we expect to find several supernovae (SNe) exploding in
background galaxies. There are eleven such findings. Two of them are most
likely SN Ia and the remaining nine are also good SN candidates.

\begin{figure}[t]
\centerline{\includegraphics[width=12.3cm]{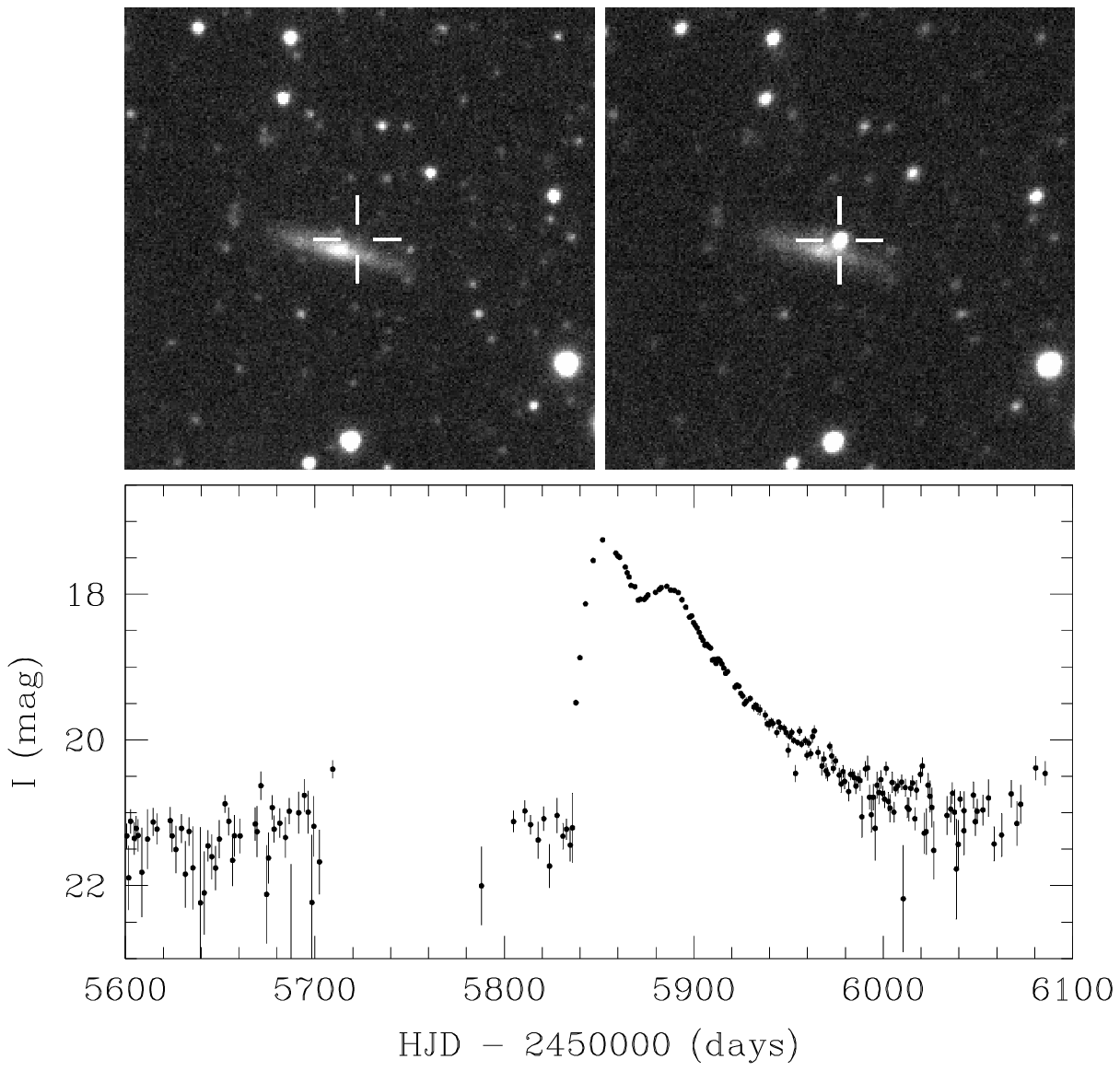}}
\vspace*{3mm}
\FigCap{{\it Top:} Finding chart for supernova OGLE-2011-SN-003. 
{\it Left image} shows the galaxy before SN explosion, while the {\it right
image} shows the SN at its peak (marked with cross hairs). It is located
2\zdot\arcs7 away from a nearly edge-on galaxy
OGLE-GALAXY-LMC563.17.14. The image covers $60\arcs\times60\arcs$. {\it
Bottom:} \mbox{OGLE-IV} light curve for SN OGLE-2011-SN-003. It peaked at
$I=17.27$ mag on 2011, October 17. The gap at $5700<{\rm HJD}-2450000<5800$
is the seasonal gap, when the LMC is not observable.}
\end{figure}

\begin{figure}[htb]
\centerline{\includegraphics[width=12.3cm]{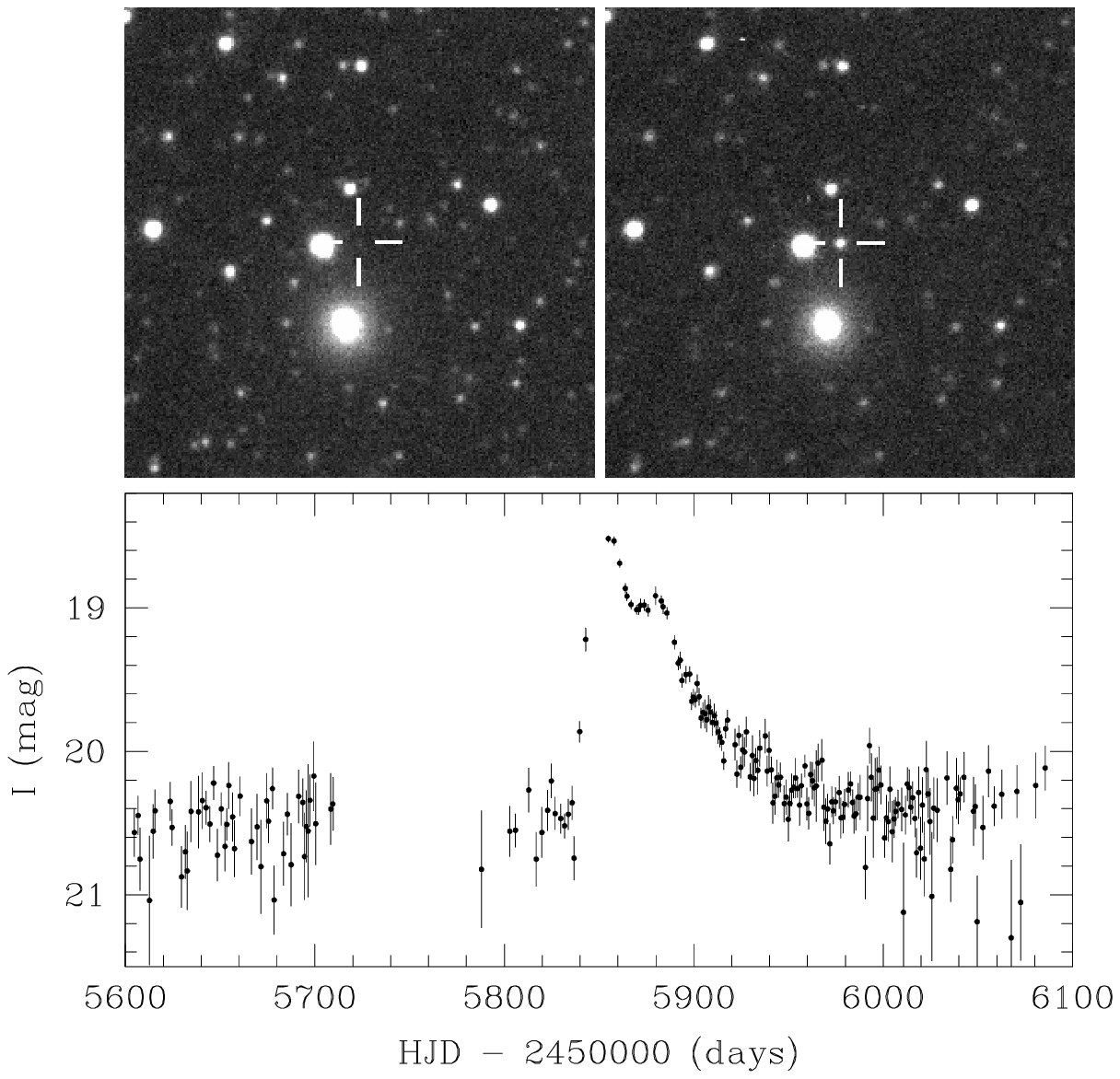}}
\vspace*{3mm}
\FigCap{{\it Top:} Finding chart for supernova OGLE-2011-SN-004. 
{\it Left image} shows the galaxy before SN explosion, while the {\it right
image} shows the SN at its peak (marked with cross hairs). It is located
10\zdot\arcs7 away from an elliptical galaxy OGLE-GALAXY-LMC570.28.8. The
image covers $60\arcs\times60\arcs$. {\it Bottom:} \mbox{OGLE-IV} light
curve for SN OGLE-2011-SN-004. It peaked at $I=18.51$~mag on 2011, October
20.}
\end{figure}

The supernova OGLE-2011-SN-003 (Fig.~13) appeared 2\zdot\arcs7 away from a
nearly edge-on galaxy OGLE-GALAXY-LMC563.17.14 (see Section~5.2 describing
the galaxy catalog). The second object -- supernova OGLE-2011-SN-004,
(Fig.~14) was detected 10\zdot\arcs7 away from an elliptical galaxy
OGLE-GALAXY-LMC570.28.8. Both these objects were found during the search of
the databases of new objects. An artificial blend star of $I=20.65$~mag was
added to the light of both objects to avoid undefined magnitudes when the
stars were not detectable on the frames before and after brightening.

The additional nine objects were found during the search of
regular \mbox{OGLE-IV} databases. Faint galaxies often mimic faint stars
and therefore they are included to these databases. Explosion of a SN in
such a galaxy can be recognized by a characteristic spike on the otherwise
flat and noisy galaxy light curve. Additional visual confirmation on the
reference image that the object is indeed a galaxy makes the SN
interpretation of a transient very sound.

\begin{figure}[htb]
\centerline{\includegraphics[width=12.3cm]{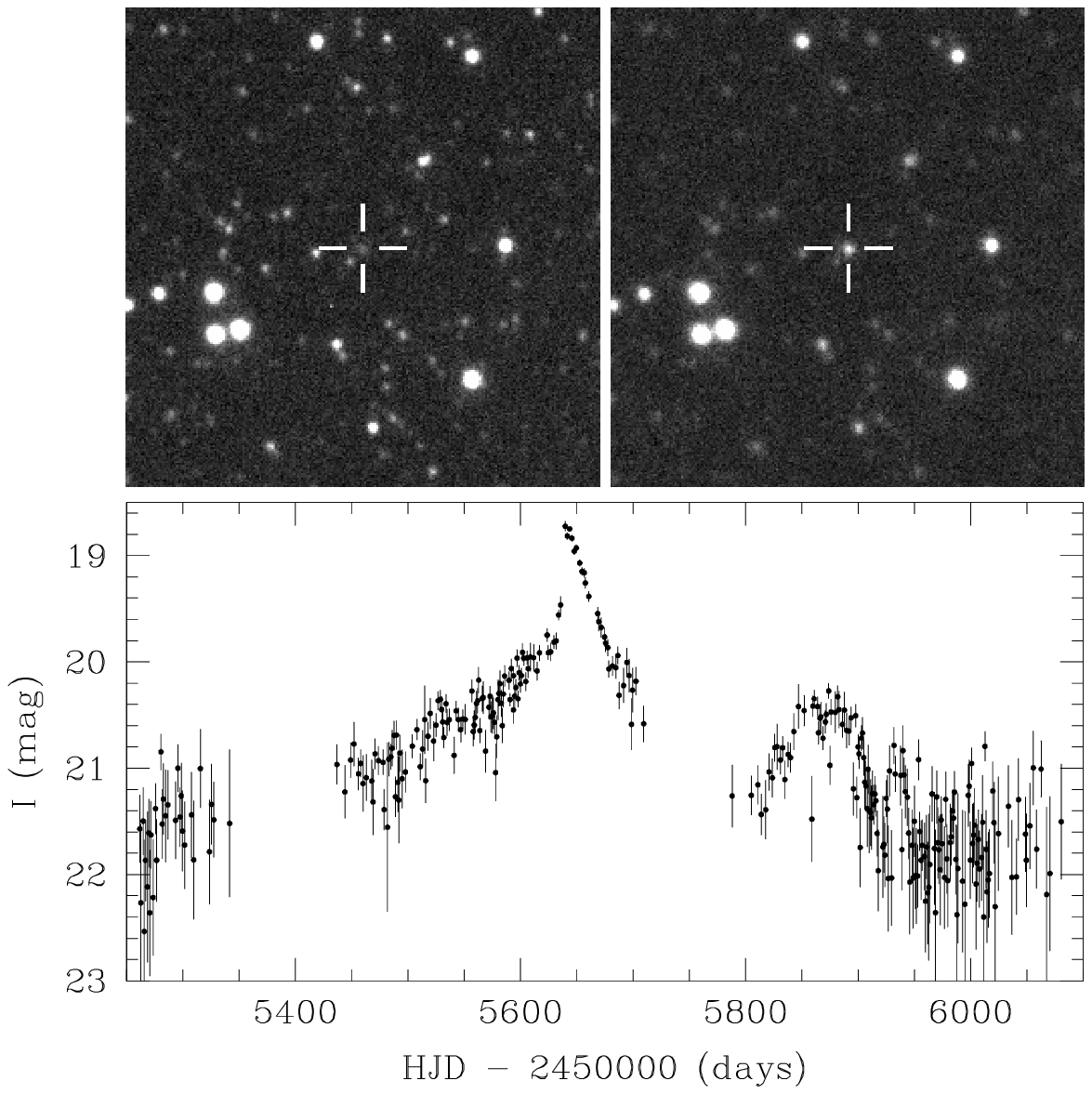}}
\vspace*{3mm}
\FigCap{{\it Top:} Finding chart for supernova candidate,
OGLE-2011-SN-001. {\it Left image} shows the galaxy before SN
explosion, while the {\it right image} shows the SN at its peak (marked
with cross hairs). {\it Bottom:} \mbox{OGLE-IV} light curve for SN candidate
OGLE-2011-SN-001. It peaked at $I=18.74$~mag on 2011, March 19.}
\end{figure}

An example of an interesting SN candidate found in this search,
OGLE-2011-SN-001, is presented in Fig.~15. The light curve clearly shows a
supernova-type spike superimposed on the long term brightening. Next
observing season, after the fading of the main supernova-like peak -- the
second brightening episode clearly occurred. Evidently, this object is
located on the faint galaxy, therefore it is likely a supernova candidate.

\MakeTableee{c@{\hspace{5pt}}l@{\hspace{5pt}}c@{\hspace{5pt}}c@{\hspace{5pt}}c@{\hspace{5pt}}c@{\hspace{0pt}}c}{12.5cm}{SN candidates in the GSEP Field}
{\hline
\noalign{\vskip3pt}
ID & \multicolumn{1}{c}{\mbox{OGLE-IV}} &    R.A.    &   Dec    & $I_{\rm max}$ & $T_{\rm max}$    & {\scriptsize Remarks}\\
   &\multicolumn{1}{c}{Object ID} & [2000.0] & [2000.0] & [mag]         & JD [days]         & \\
\noalign{\vskip3pt}
\hline
\noalign{\vskip3pt}
OGLE-2010-SN-001 & LMC571.04.7642   & 06\uph04\upm14\zdot\ups54 & $-66\arcd00\arcm24\zdot\arcs3$ & 19.73 & $2\;455\;545$ & G1 \\
OGLE-2011-SN-001 & LMC563.10.12139  & 05\uph57\upm27\zdot\ups81 & $-66\arcd21\arcm42\zdot\arcs6$ & 18.74 & $2\;455\;640$ & G \\ 
OGLE-2011-SN-002 & LMC571.23.5674   & 06\uph01\upm00\zdot\ups48 & $-65\arcd25\arcm51\zdot\arcs9$ & 20.15 & $2\;455\;813$ & G2 \\
OGLE-2011-SN-003 & LMC563.17.499N   & 05\uph59\upm50\zdot\ups14 & $-65\arcd56\arcm22\zdot\arcs8$ & 17.27 & $2\;455\;852$ & G3 \\ 
OGLE-2011-SN-004 & LMC570.28.179N   & 06\uph07\upm16\zdot\ups62 & $-66\arcd20\arcm24\zdot\arcs0$ & 18.51 & $2\;455\;855$ & G4 \\ 
OGLE-2011-SN-005 & LMC571.28.4728   & 06\uph05\upm24\zdot\ups89 & $-65\arcd03\arcm03\zdot\arcs1$ & 19.79 & $2\;455\;858$ & G5 \\
OGLE-2011-SN-006 & LMC570.31.2207   & 06\uph02\upm12\zdot\ups93 & $-66\arcd27\arcm45\zdot\arcs8$ & 19.83 & $2\;455\;913$ &  \\
OGLE-2012-SN-001 & LMC563.32.5268   & 05\uph49\upm45\zdot\ups04 & $-65\arcd47\arcm46\zdot\arcs9$ & 19.73 & $2\;455\;940$ & G \\ 
OGLE-2012-SN-002 & LMC571.04.3566   & 06\uph04\upm34\zdot\ups87 & $-66\arcd06\arcm13\zdot\arcs6$ & 19.59 & $2\;455\;956$ & G \\
OGLE-2012-SN-003 & LMC570.12.8488   & 06\uph06\upm30\zdot\ups17 & $-66\arcd57\arcm04\zdot\arcs2$ & 19.53 & $2\;455\;992$ & G \\
OGLE-2012-SN-004 & LMC571.19.642    & 06\uph06\upm44\zdot\ups62 & $-65\arcd35\arcm49\zdot\arcs4$ & 19.43 & $2\;456\;007$ & G \\
\noalign{\vskip3pt}
\hline
\noalign{\vskip3pt}
\multicolumn{7}{p{12.5cm}}{Notes: G -- SN host galaxy is present on the template image,}\\
\multicolumn{7}{p{12.5cm}}{G1 -- galaxy OGLE-GALAXY-LMC571.04.11,}\\
\multicolumn{7}{p{12.5cm}}{G2 -- galaxy OGLE-GALAXY-LMC571.23.12,}\\
\multicolumn{7}{p{12.5cm}}{G3 -- galaxy OGLE-GALAXY-LMC563.17.14,}\\
\multicolumn{7}{p{12.5cm}}{G4 -- galaxy OGLE-GALAXY-LMC570.28.8,}\\
\multicolumn{7}{p{12.5cm}}{G5 -- galaxy OGLE-GALAXY-LMC571.28.22.}
}

Table~3 lists all the SN candidates detected in the \mbox{OGLE-IV} GSEP field.
Their light curves are available in the OGLE Internet archive (see
Section~7).

The detection of eleven SNe candidates in the GSEP field over 2.2 years
indicates the rate of SN occurrence of about 2~SN per year per square
degree including the seasonal gaps (20\% of time) in observations and
assuming about 70--80\% detection efficiency. This is a similar rate to
that found during the search for SNe in the neighboring Magellanic Bridge
fields also observed by \mbox{OGLE-IV} for similar period of time
(Koz³owski \etal 2012, in preparation).

\Subsection{Catalog of Galaxies}

Identification and confirmation of supernovae, one of the main targets of
the Gaia alerting pipeline, will benefit strongly if the possible supernova
was cross-matched with a nearby galaxy. \mbox{OGLE-IV} images of the GSEP
region provide an excellent material for cataloging most of the galaxies up
to redshift $z\approx0.1$, which is the expected limit for supernovae to be
detected by Gaia.

\begin{figure}[t]
\vglue-0.7cm
\centerline{\includegraphics[width=13cm]{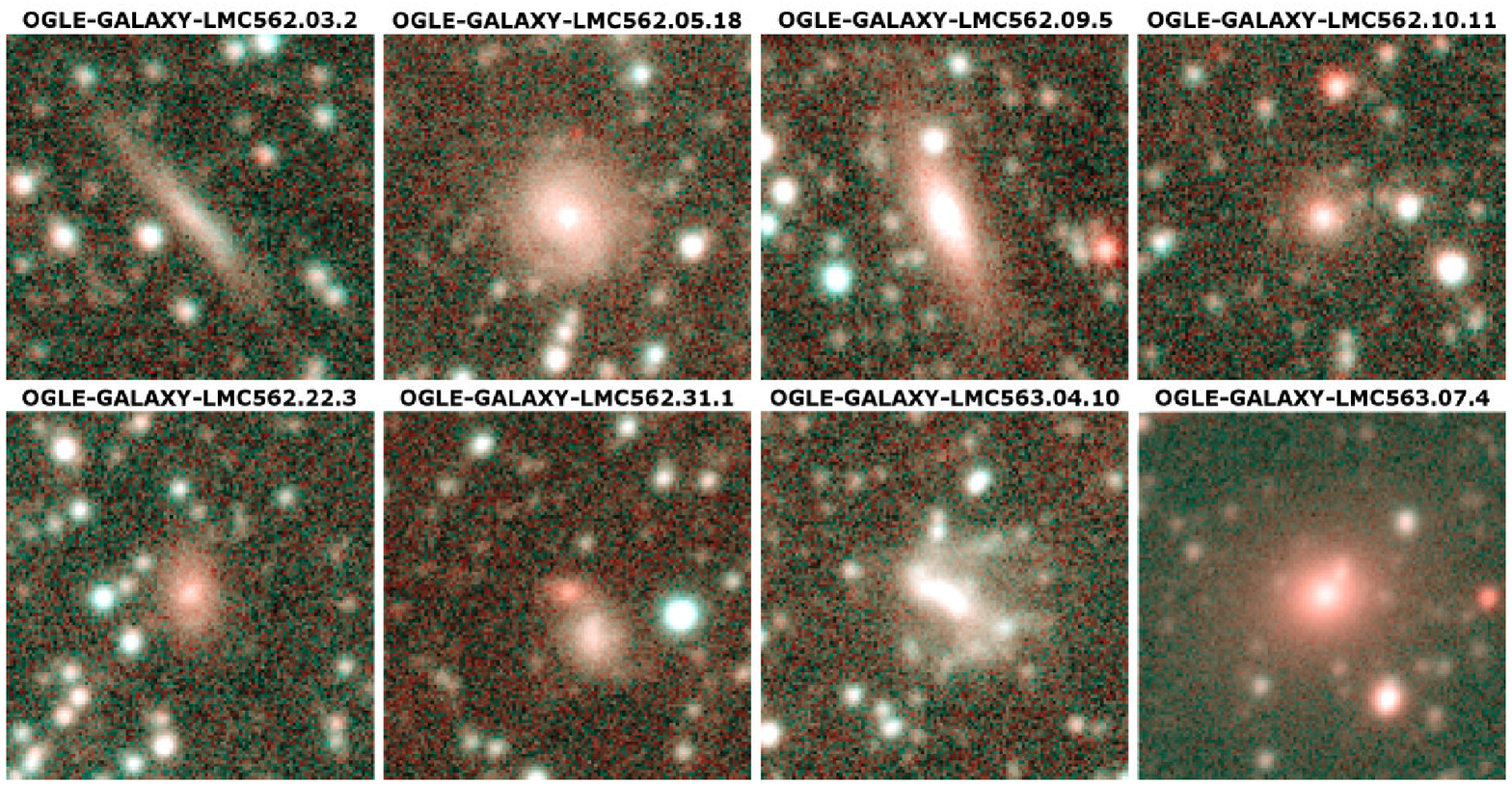}}
\vspace*{1mm}
\FigCap{Examples of galaxies identified in the \mbox{OGLE-IV} reference 
images of the GSEP field.}
\end{figure}

We searched for galaxies in the \mbox{OGLE-IV} GSEP field images using {\sc
SExtractor} (Bertin and Arnouts 1996) running on the DIA reference {\it
I}-band images. We selected as potential galaxies all objects brighter than
17.5~mag with class\_star~<~0.15 and brighter than 18.7~mag with
class\_star~<~0.01. Then, we visually inspected all of about 4000
candidates to select 1925 apparent galaxies. Fig.~16 shows examples of
identified galaxies as a false-color composition of {\it I}- and {\it
V}-band images. Fig.~17 shows the locations of all identified galaxies on
the map of \mbox{OGLE-IV} GSEP field, with red and blue points representing
galaxies brighter and fainter, respectively, than 18~mag in the {\it
I}-band. Table containing all 1925 candidate galaxies, along with their
coordinates and {\it I}-band magnitudes is available from the OGLE Internet
archive (see Section~7).

\begin{figure}[t]
\vglue-9mm
\centerline{\includegraphics[width=13.3cm]{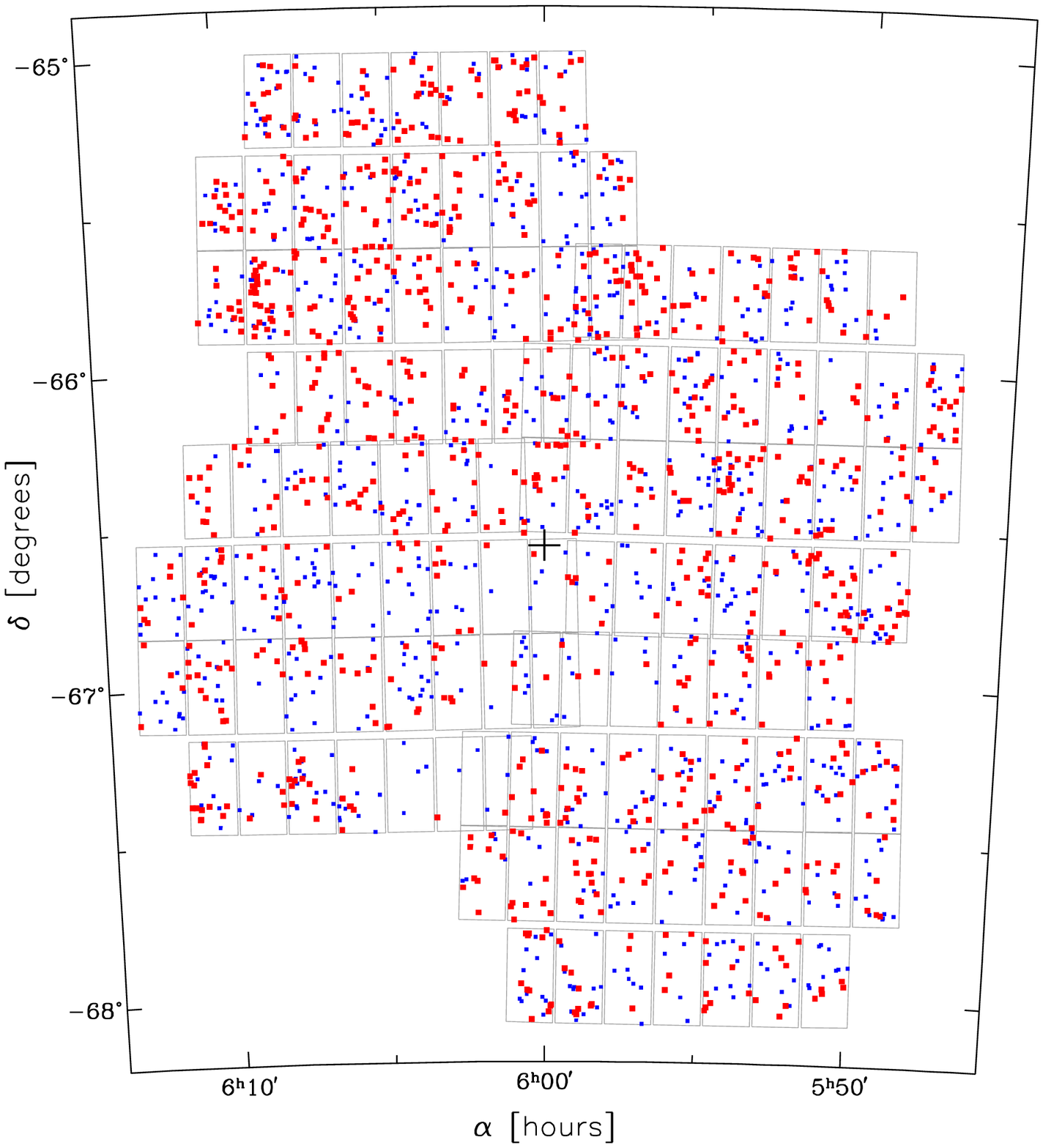}}
\vspace*{-30mm}
\FigCap{Spatial distribution of galaxies detected in the GSEP field.
Gray contours show the sky coverage by the individual chips of the
\mbox{OGLE-IV} mosaic CCD camera. Red and blue symbols show positions of
galaxies brighter and fainter, respectively, than 18~mag in the {\it
I}-band. Black cross indicates the position of the South Ecliptic Pole.}
\end{figure}

\vspace*{-7pt}
\Section{Astrometry of the GSEP Field}
\vspace*{-5pt}
The OGLE images collected at one of the best observational sites worldwide,
Las Campanas Observatory, with high angular resolution of the OGLE cameras
and long time span are ideal observational material for astrometric
purposes (Poleski \etal 2012). Therefore, we additionally analyzed the
time-series astrometry of all stars brighter than $I=19.5$~mag in the GSEP
field. The centroids of the stellar images were measured for each epoch
separately using our own code which is based on method presented by
Anderson and King (2000). The proper motions and parallaxes were derived
similarly to Poleski \etal (2012) and are tied to the LMC stars. The
typical accuracy of the derived proper motions is 1.8~mas/yr per
coordinate, while typical accuracy of the parallaxes is 1.5~mas. We failed
in analyzing two subfields namely LMC570.17 and LMC571.17.

In total, we derived reliable proper motions for 3309 stars from the GSEP
field. To avoid artifacts we only selected objects with $\chi^2$ of the
astrometric fit smaller than 2.5, positive cross-match in at least 70\% of
well registered epochs of the given subfield and proper motion greater than
20~mas/yr. The stars with proper motions greater that 50~mas/yr were
visually verified. For the stars with proper motions between 20~mas/yr and
50~mas/yr the false positive rate is 10\% and the list is complete in 90\%.

Significant proper motions were measured for 50 variable stars, which
clearly indicates that these stars belong to the Milky Way. We also found
50 non-variable high proper motion (HPM) stars, \ie the ones with the
proper motion larger than 100~mas/yr or slightly below that limit, but with
parallaxes greater than 10~mas. Based on measured parallaxes, {\it I}-band
magnitudes and $(V-I)$ colors we found three of the HPM stars to be nearby
white dwarfs.

The file {\sf pm.dat} in the OGLE Internet archive lists all objects with
reliable astrometry found in the GSEP field. For each star, apart from the
\mbox{OGLE-IV} identification number, coordinates and mean brightness in
the {\it I}- and {\it V}-bands, we present the proper motion in right
ascension and declination with uncertainties.

Additionally, the file {\sf hpm.dat} contains the results of our search for
HPM stars. Here, in addition to the information provided in {\sf pm.dat}
file the equatorial coordinates of the object position at the J2000 epoch
are given. Please note that the position of some of the stars listed in
this file changed up to 4\arcs on \mbox{OGLE-IV} images taken more than ten
years after the J2000 epoch. Also, for all objects the parallax and its
uncertainty are given. Three white dwarfs found are marked.

One has to remember that the \mbox{OGLE-IV} astrometry is based on
relatively short time baseline of only 26 months. The data presented here
will be superseded in the future by the next releases of the OGLE
astrometric catalogs.

\Section{Data Availability}
All data presented in this paper are available to the astronomical
community from the OGLE Internet archive accessible from the OGLE WWW Page
or directly:

\begin{center}
{\it http://ogle.astrouw.edu.pl}\\
{\it ftp://ftp.astrouw.edu.pl/ogle/ogle4/GSEP}
\end{center}

Please read the {\sf README} file for the details on the data presented
there as well as on all updates.

\Acknow{We would like to thank Z.~Ko³aczkowski and
A.~Schwar\-zenberg-Czerny for providing software which enabled us to
prepare this study. We would also like to thank G.~Clementini and
L.~Eyer for stimulating discussions about this project.

The OGLE project has received funding from the European Research Council
under the European Community's Seventh Framework Programme
(FP7/2007-2013)/ERC grant agreement no. 246678 to AU. This work has been
supported by the Polish National Science Centre grant no.
DEC-2011/03/B/ST9/02573. The astrometric part of the research was
supported by Polish Ministry of Science and Higher Education through the
program ''Iuventus Plus'' award No. IP2011 043571 to RP.}


\begin{references}
\refitem{Alard, C., and Lupton, R.H.}{1998}{\ApJ}{503}{325}
\refitem{Anderson, J., and King, I.R.}{2000}{\PASP}{112}{1360}
\refitem{Artyukhina, N.M., \etal}{1995}{~}{~}{General Catalogue of Variable Stars, 4rd ed., vol.V. Extragalactic Variable Stars, "Kosmosinform", Moscow}
\refitem{Bertin, E., and Arnouts, S.}{1996}{\AAS}{117}{393}
\refitem{Cioni, M.-R.L., Clementini, G., Girardi, L., \etal}{2011}{\AA}{527}{A116}
\refitem{Cutri, R.M., \etal}{2003}{~}{~}{``2MASS All-Sky Catalog of Point Sources''}
\refitem{de Bruijne, J.H.J.}{2012}{\it Ap\&SS}{341}{31}
\refitem{Leavitt, H.S.}{1908}{Ann. Harv. Coll. Obs.}{60}{87}
\refitem{Mennickent, R.E., Pietrzy{\'n}ski, G., Diaz, M., and Gieren, W.}{2003}{\AA}{399}{L47}
\refitem{Pietrzyñski, G., Thompson, I.B., Gieren, W., Graczyk, D., Bono, G., Udalski, A., Soszyñski, I., Minniti, D., and Pilecki, B.}{2010}{Nature}{468}{542}
\refitem{Pietrzyñski, G., Thompson, I.B., Graczyk, D., Gieren, W., Pilecki, B., Udalski, A., Soszyñski, I., Bono, G., Konorski, P., Nardetto, N., and Storm, J.}{2011}{\ApJ}{742}{L20}
\refitem{Pigulski, A., Ko³aczkowski, Z., Ramza, T., and Narwid, A.}{2006}{Mem. Soc. Astron. Ital.}{77}{223}
\refitem{Pojmañski, G.}{1997}{\Acta}{47}{467}
\refitem{Poleski,~R., Soszyñski,~I., Udalski, A., Szymañski,~M.K., Kubiak,~M., Pietrzyñski,~G., Wyrzykowski,~£., Szewczyk,~O., and Ulaczyk,~K.}{2010a}{\Acta}{60}{1}
\refitem{Poleski,~R., Soszyñski,~I., Udalski, A., Szymañski,~M.K., Kubiak,~M., Pietrzyñski,~G., Wyrzykowski,~£., and Ulaczyk,~K.}{2010b}{\Acta}{60}{179}
\refitem{Poleski,~R., Soszyñski,~I., Udalski, A., Szymañski,~M.K., Kubiak,~M., Pietrzyñski,~G., Wyrzykowski,~£., and Ulaczyk,~K.}{2012}{\Acta}{62}{1}
\refitem{Schwarzenberg-Czerny, A.}{1996}{\ApJ}{460}{L107}
\refitem{Shapley, H., and Mohr, J.}{1933}{Ann. Harv. Col. Obs.}{90}{1}
\refitem{Soszyñski, I., Dziembowski, W.A., Udalski, A., Kubiak, M., Szymañski, M.K., Pietrzyñski, G., Wyrzykowski, {\L}., Szewczyk,~O., and Ulaczyk, K.}{2007}{\Acta}{57}{201}
\refitem{Soszyñski,~I., Poleski,~R., Udalski, A., Szymañski,~M.K., Kubiak,~M., Pietrzyñski,~G., Wyrzykowski,~£., Szewczyk,~O., and Ulaczyk,~K.}{2008}{\Acta}{58}{163}
\refitem{Soszyñski,~I., Udalski, A., Szymañski,~M.K., Kubiak,~M., Pietrzyñski,~G., Wyrzykowski,~£., Szewczyk,~O., Ulaczyk,~K., and Poleski,~R.}{2009a}{\Acta}{59}{1}
\refitem{Soszyñski,~I., Udalski, A., Szymañski,~M.K., Kubiak,~M., Pietrzyñski,~G., Wyrzykowski,~£., Szewczyk,~O., Ulaczyk,~K., and Poleski,~R.}{2009b}{\Acta}{59}{239}
\refitem{Szymañski, M.K., Udalski, A., Soszyñski, I., Kubiak, M., Pietrzyñski, G., Poleski, R., Wyrzykowski,~£., and Ulaczyk, K.}{2011}{\Acta}{61}{83}
\refitem{Tisserand, P., Le Guillou, L., Afonso, C., \etal}{2007}{\AA}{469}{387} 
\refitem{Udalski, A.}{2003}{\Acta}{53}{291}
\refitem{Udalski, A., Szyma{\'n}ski, M.K., Soszy{\'n}ski, and Poleski, R.}{2008}{\Acta}{58}{69}
\refitem{Wetzel, M.A.}{1955}{Ann. Harv. Col. Obs.}{109}{58}
\refitem{Wood, P.R., \etal (MACHO team)}{1999}{~}{~}{in: {\it IAU Symp.} 191, ``Asymptotic Giant Branch Stars'', ed. T.~Le~Bertre, A. L\'ebre, and C. Waelkens (San Francisco: ASP), p.~151}
\refitem{Wo¼niak, P.R.}{2000}{\Acta}{50}{421}
\refitem{Wyrzykowski, £., and Hodgkin, S.}{2012}{~}{~}{in IAU Symp. 285, New Horizons in Time-Domain Astronomy, ed. R. E. M. Griffin, R. J. Hanisch, and R. Seaman (Cambridge: Cambridge Univ. Press), 425}
\end{references}
\end{document}